\documentclass[aps,amsmath,preprint,superscriptaddress,nofootinbib]{revtex4-1}  

\usepackage{graphicx}
\usepackage{color}
\usepackage{amsmath,amssymb}	
\usepackage{hyperref}
\usepackage{setspace}
\usepackage{array}  
\usepackage{cancel}
\usepackage{enumitem}
\usepackage{mathrsfs}
\usepackage{physics}
\usepackage{ulem}

\newcommand{\xx}{\boldsymbol{x}}

\begin{document}
\newcolumntype{Y}{>{\centering\arraybackslash}p{23pt}} 


\preprint{IPMU20-0059, RUP-20-16}

\title{Cosmic String in Abelian-Higgs Model with Enhanced Symmetry -- Implication to the Axion Domain-Wall Problem --}

\author{Takashi Hiramatsu}
\email[e-mail: ]{hiramatz@rikkyo.ac.jp}
\affiliation{ICRR, The University of Tokyo, Kashiwa, Chiba 277-8582, Japan}
\affiliation{Department of Physics, Rikkyo University, Toshima, Tokyo 171-8501, Japan}
\author{Masahiro Ibe}
\email[e-mail: ]{ibe@icrr.u-tokyo.ac.jp}
\affiliation{ICRR, The University of Tokyo, Kashiwa, Chiba 277-8582, Japan}
\affiliation{Kavli IPMU (WPI), UTIAS, The University of Tokyo, Kashiwa, Chiba 277-8583, Japan}
\author{Motoo Suzuki}
\email[e-mail: ]{m0t@icrr.u-tokyo.ac.jp}
\affiliation{Tsung-Dao Lee Institute, Shanghai Jiao Tong University, Shanghai 200240, China}
\date{\today}
\begin{abstract}
In our previous work, we found new types of the cosmic string solutions in the Abelian-Higgs model with an enhanced $U(1)$ global symmetry.
We dubbed those solutions as the compensated/uncompensated strings. 
The compensated string is similar to the conventional cosmic string in the Abrikosov-Nielsen-Olesen (ANO) string, 
around which only the would-be Nambu-Goldstone (NG) boson winds. Around the uncompensated string, on the other hand, the physical NG boson also winds, where 
the physical NG boson is associated with the spontaneous breaking of the enhanced symmetry.
Our previous simulation in the 2+1 dimensional spacetime confirmed that both the compensated/uncompensated strings are formed at the phase transition of the symmetry breaking.
Non-trivial winding of the physical NG boson around the strings potentially causes the so-called axion domain-wall problem when the model is applied to the axion model.
In this paper, we perform simulation in the 3+1 dimensional spacetime to discuss the fate of the uncompensated strings.
We observe that the evolution of the string-network is highly complicated in the 3+1 dimensional simulation compared with that seen in the previous simulation.
Despite such complications, 
we find that the number of the uncompensated strings which could cause can be highly suppressed at late times.
Our observation suggests that the present setup can be applied to the axion model without suffering from the axion domain-wall problem.
\end{abstract}
\maketitle
\newpage
\section{Introduction}
\label{sec:introduction}
Recently, we discussed new types of string (vortex) solutions in the Abelian-Higgs model with two complex scalar fields~\cite{Hiramatsu:2019tua}.
As a peculiar feature, the model has
an accidental $U(1)$ global symmetry, $U(1)_\mathrm{global}$, enhanced by the hierarchical charge assignment of the $U(1)$ gauge symmetry, $U(1)_\mathrm{local}$.
By the spontaneous breaking of $U(1)_\mathrm{local}$ and $U(1)_\mathrm{global}$, one of the Nambu-Goldstone (NG) boson is absorbed by the Higgs mechanism, while the other one appears as a physical NG boson. 
This class of setup has been discussed in the context of the QCD axion models where a global Peccei-Quinn symmetry~\cite{Peccei:1977hh,Peccei:1977ur,Weinberg:1977ma,Wilczek:1977pj} is
identified with $U(1)_\mathrm{global}$~\cite{Barr:1982uj,Choi:1985iv,Fukuda:2017ylt,Fukuda:2018oco,Ibe:2018hir}. In those applications, the gauge symmetry protects the PQ symmetry from the explicit breaking caused by the quantum gravity effects~\cite{Hawking:1987mz,Lavrelashvili:1987jg,Giddings:1988cx,Coleman:1988tj,Gilbert:1989nq,Banks:2010zn}.

In \cite{Hiramatsu:2019tua}, we call the new solutions, the ``compensated" and the ``uncompensated" strings.
The compensated string is more like the local string~\cite{Nielsen:1973cs},
in which only the would-be NG boson winds non-trivially.
The uncompensated string has, on the other hand, properties in between the local string and the global string solutions~\cite{Nielsen:1973cs}. 
Around the uncompensated strings, not only the would-be NG boson but also the physical NG boson wind.%
\footnote{Similar to the compensated/uncompensated string, the composite string solutions in the de Sitter space have been discussed in~\cite{BezerradeMello:2003ei}.}
We also found that the uncompensated strings have long-range repulsive and attractive forces.%
\footnote{This type of the string solutions has been discussed in~\cite{Hill:1987bw}, where the evolution of the string network is considered 
in the presence of the explicit breaking of the global $U(1)$ symmetry.
In this paper, we discuss how the network evolves in the absence of the explicit breaking, 
and hence, no domain walls are attached until we apply the model to the axion model.
We thank Guy Moore for letting us know this previous work.}

In our previous work~\cite{Hiramatsu:2019tua}, we also performed
classical lattice simulation of the time-evolution in the spacetime of the 2+1 dimension (2+1\,D). (The earlier numerical simulation of the vortex formation in a similar setup has been performed in~\cite{Garaud:2014laa}, which exhibits consistent features with our simulations in~\cite{Hiramatsu:2019tua}.) The simulation confirmed the ubiquitous formation of the uncompensated strings
at the phase transition.
As a remarkable feature, the compensated/uncompensated strings have multiple winding numbers.
This should be contrasted with the string formation in the conventional Abelian-Higgs model,
where the strings with multiple winding numbers are hardly formed
at the phase transition.
We also found a tendency that the uncompensated strings evolve into the local strings by the long-range forces mentioned above. Indeed, at a later time, we observed that most of the uncompensated strings end up being the local string in the 2+1\,D simulation.

In this paper, we explore the formation and the evolution of the string network by performing 3+1\,D classical lattice simulation (see the pioneer works using lattice simulations for the Abelian-Higgs strings and the global strings, e.g., \cite{Vincent:1997cx,Moore:2001px,Salmi:2007ah,Achucarro:2006es,Yamaguchi:1998gx,Yamaguchi:1999dy,Yamaguchi:1999yp,Yamaguchi:2002zv,Yamaguchi:2002sh,Hiramatsu:2013tga,Higaki:2016jjh,Hindmarsh:2018wkp}).
As we will see, the collision and the reconnection of the uncompensated strings are much more complicated 
than those seen in 
the 2+1\,D simulation.%
\footnote{In the 2+1\,D simulation, the strings appear as ``particles" in the 2\,D plane, and hence, the reconnection just looks like the binding of two particles. }
In particular, we observe the formation of  ``bridges'' between two strings, which is a peculiar feature of the present model.
The bridge formation makes the cosmological evolution of the string network highly complicated compared with the conventional string networks.
The formation of the bridges also makes the combination process 
of the uncompensated strings into the compensated strings less efficient in the 3+1\,D than that in the 2+1\,D.
The uncompensated strings potentially cause the axion domain-wall problem 
when this model is applied to the QCD axion.
Thus, the 3+1\,D simulation is imperative for the axion domain-wall problem where the PQ symmetry is identified with $U(1)_\mathrm{global}$.%
\footnote{See \cite{Chatterjee:2019rch} for the string-wall network in the QCD axion model appearing in the non-Abelian gauge symmetry~\cite{Lazarides:1982tw}.}

The organization of the paper is as follows. In Sec.~\ref{sec:review}, we review the setup of the model and the results in our previous work. In Sec.~\ref{sec:collision}, we simulate the collision of two strings and show the results. In Sec.~\ref{sec:simulation}, we describe the setup and analysis to study the evolution of the string networks. In Sec.~\ref{sec:result}, we perform simulation of the time-evolution of the string network. In Sec.~\ref{sec:discussion}, we discuss the implications to the axion domain-wall problem.
The final section is devoted to the summary.

\section{Review of New String Solutions in Abelian-Higgs Model}
\label{sec:review}
In this section, we summarize the model setup. We also review the compensated/uncompensated string solutions, as well as the results of the simulation in the 2+1\,D spacetime~\cite{Hiramatsu:2019tua}.

\subsection{Model}

The action of the model is given by,%
\footnote{The metric $(-,+,+,+)$ is employed in this paper.}
%
\begin{align}
&S=-\int d^4x\,\left[(\mathscr{D}_\mu\phi_1)^*\mathscr{D}^\mu\phi_1
                + (\mathscr{D}_\mu\phi_2)^*\mathscr{D}^\mu\phi_2
                +V(\phi_1,\phi_2)+\frac{1}{4}F_{\mu\nu}F^{\mu\nu}\right]\ ,\\
&\mathscr{D}_\mu\phi_n=\partial_\mu\phi_n-ieq_nA_\mu \phi_n\ ,~F_{\mu\nu}=\partial_\mu A_\nu-\partial_\nu A_\mu\ .             
\end{align}
%
Here, $\phi_n$ $(n=1,2)$ are the two complex scalar fields with $U(1)_\mathrm{local}$ charge $q_{n}$, $A_\mu$ is the $U(1)_\mathrm{local}$ gauge field, and $e$ denotes the gauge coupling constant of $U(1)_\mathrm{local}$. We take $q_1$ and $q_2$ to be relatively prime numbers without loss of generality. The scalar potential $V(\phi_1,\phi_2)$ is given by
%
\begin{align}
\label{eq:potential}
V(\phi_1,\phi_2)=\frac{\lambda_1}{4}(|\phi_1|^2-\eta_1^2)^2+\frac{\lambda_2}{4}(|\phi_2|^2-\eta_2^2)^2-\kappa(|\phi_1|^2-\eta_1^2)(|\phi_2|^2-\eta_2^2)\ ,
\end{align}
%
where $\lambda_{1,2}$ and $\kappa$ are real dimensionless constants, and $\eta_{1,2}$ are real constants with mass dimension one. 
The model possesses an $U(1)_\mathrm{global}$ symmetry in addition to the $U(1)_\mathrm{local}$ symmetry.
Such $U(1)_\mathrm{global}$ symmetry naturally appears as an accidental symmetry in the renormalizable theory when $|q_1| + |q_2|$ is larger than $4$.
We assume $\lambda_1 \lambda_2 > 4 \kappa^2$ so that both the $\phi_{1,2}$ obtain non-vanishing vacuum expectation values (VEV's),
\begin{align}
\label{eq:vacuum}
    \langle \phi_n \rangle = \eta_n \ , \quad (n = 1,2)\ .
\end{align}

The VEV's in Eq.\,\eqref{eq:vacuum}
break 
both the $U(1)_\mathrm{local}$ and $U(1)_\mathrm{global}$ symmetries spontaneously.
Accordingly,
there appear two NG modes, where one corresponds to the would-be NG boson $(b)$, and another one is the gauge-invariant NG modes $(a)$. 
We can extract these two modes from the phase components of the complex scalar fields,
%
\begin{align}
\label{eq:axialcomp}
&\phi_1 = \frac{1}{\sqrt{2}}f_1\, e^{i \tilde a_1/f_1}\ , \quad 
\phi_2 =\frac{1}{\sqrt{2}}f_2\, e^{i \tilde a_2/f_2}\ .
\end{align}
%
Here,  $\tilde a_{1,2}$ are pseudo-scalar fields,
and $f_n\equiv \sqrt{2}\eta_n$ $(n=1,2)$ are the decay constants of them.
The gauge-invariant and the would-be NG modes, $a$ and $b$, are given by~\cite{Fukuda:2017ylt,Fukuda:2018oco},
%
\begin{align}
\left(
\begin{array}{cc}
 a   \\
 b
\end{array}
\label{eq:decomp}
\right)=
\frac{1}{\sqrt{q_{1}^2f^{2}_1 
+q_{2}^2f^{2}_2 }}\left(
\begin{array}{cc}
q_{2} f_2   &  -q_{1} f_1   \\
q_{1} f_1  &   q_{2} f_2
\end{array}
\right)
\left(
\begin{array}{cc}
 \tilde a_1   \\
\tilde a_2
\end{array}
\right)\ .
\end{align}
%
We can see the component $a$ is indeed invariant under the $U(1)_\mathrm{local}$ gauge transformation,
%
\begin{align}
\label{eq:shift}
\tilde{a}_{1}/f_1\to\tilde{a}_{1}/
f_1 + q_{1} \alpha \ ,
\quad \tilde{a}_{2}/f_2  \to\tilde{a}_{2}/f_2 + q_{2} \alpha \ ,
\end{align}
%
where $\alpha$ denotes the $U(1)_\mathrm{local}$ gauge transformation parameter. 
The domain of the gauge invariant axion is given by,
%
\begin{align}
\label{eq:domain}
\frac{a}{F_a} \in [0,2\pi) \ , \quad F_a \equiv \frac{f_1f_2}
    {\sqrt{q^{2}_1 f^{2}_1 + 
    q^{2}_2f^{2}_2 }}\ .
\end{align}
%

\subsection{Uncompensated String}
At the phase transition, we expect string formation due to the non-trivial
homotopy group of the vacuum manifold, $\pi_1(U(1)_\mathrm{local}\times U(1)_\mathrm{global}) = \mathbb{Z}\times \mathbb{Z}$.
In fact, we can obtain the static string solutions under the ansatz,
%
\begin{align}
\label{eq:string1}
\phi_1(r,\theta)&=\eta_1e^{in_1\theta}h_1(r)\ ,\\
\label{eq:string2}
\phi_2(r,\theta)&=\eta_2e^{in_2\theta}h_2(r)\ ,\\
\label{eq:gauge}
A_\theta(r)&=\frac{1}{e}\xi(r)\ ,~A_r=A_z=0\ .
\end{align}
%
Here, $(r,~\theta,~z)$ are the radial distance, the azimuth angle, and the height in the cylindrical coordinate, respectively. 
The integers $n_{1}$ and $n_{2}$ 
denote the winding numbers of the strings consisting of $\phi_1$ and $\phi_2$, respectively.
The boundary conditions are
%
\begin{align}
h_1(r)=0\ ,~ h_2(r)=0\ ,~\xi(r)=0\ ,
\label{eq:boundary1}
\end{align}
 for $r \to 0$, and
%
\begin{align}
h_1(r)=1\ ,~h_2(r)=1 \ ,
\label{eq:boundary2}
\end{align} 
%
for $r\to \infty$. See~\cite{Hiramatsu:2019tua} for the numerical configurations of the static string solutions satisfying these boundary conditions.

The asymptotic behaviors of the covariant derivatives of the complex scalar fields are given by,
%
\begin{align}
\label{eq:cov1}
    \mathscr{D}_\theta \phi_1 &\to i
    \left(
    n_1 - q_{1}\frac{n_1 q_{1}\eta^{2}_1
+n_2 q_{2}\eta^{2}_2
}{q_{1}^2
\eta^{2}_1+ q_{2}^2
\eta^{2}_2} \right)
\eta_1\ ,\\
 \mathscr{D}_\theta \phi_2 &\to i
    \left(
    n_2 - q_{2}\frac{n_1 q_{1}\eta^{2}_1
+n_2 q_{2}\eta^{2}_2
}{q_{1}^2
\eta^{2}_1+ q_{2}^2
\eta^{2}_2} \right)
\eta_2\ .
\label{eq:cov2}
\end{align}
%
When they are not vanishing at $r\to \infty$, those terms lead to logarithmic divergences of the string tension at $r\to \infty$ via
%
\begin{align}
    \mu^2 \sim 2\pi \int r dr \frac{1}{r^2} |\mathscr{D}_\theta \phi_n|^2\ .
\end{align}
%
The string tension becomes finite
only when  $\mathscr{D}_\theta \phi_1$
and $\mathscr{D}_\theta \phi_2$ vanish, $i.e.$
%
\begin{align}
\label{eq:compensated}
    n_1 = N_w \times q_{1}\ , 
    \quad 
     n_2 = N_w \times q_{2}\ , \quad N_w  \in \mathbb{Z} \ .
\end{align}
%
In \cite{Hiramatsu:2019tua}, we 
call the string solutions which 
do not satisfy Eq.\,\eqref{eq:compensated} but
with $n_1 > 0$,
the uncompensated strings.
The string solutions satisfying 
Eq.\,\eqref{eq:compensated}
are the local strings which we call the compensated strings. 

As in the case of the global string,
the uncompensated string has a diverging string tension. 
Unlike the global string, however, it also has non-vanishing $U(1)_\mathrm{local}$ gauge flux at the string core. 
Therefore, the uncompensated strings have properties in between the global string and the local string.
The divergence of the string tension of the uncompensated string is cutoff by a typical distance between the strings.
Thus, it can be formed at the phase transition in the early universe, although it would have an infinite tension if it is completely isolated.

Around an uncompensated string with ($n_1$, $n_2$), not only the would-be NG boson but also the physical NG winds non-trivially
\begin{align}
& \frac{\mathit{\Delta} a}{F_a} = [0,2\pi N_{\rm dw}) \ ,  \\
&N_{\rm dw}\equiv q_2n_1-q_1n_2\ . \label{eq:def_Ndw}
\end{align}
As we will discuss in Sec.~\ref{sec:discussion}, $N_{\rm dw}$ corresponds to the domain-wall number in the application to the axion models.
Obviously, the compensated string has the vanishing domain-wall number, $N_\mathrm{dw} = 0$.

\subsection{ \texorpdfstring{2+1\,D}{Lg} Simulation Results}

In~\cite{Hiramatsu:2019tua}, we studied the formation and the evolution of the strings in the radiation dominated universe by the classical lattice simulation in the 2+1\,D spacetime. We took the parameters,
%
\begin{align}
\label{eq:parameters_d2}
(q_{1},~q_{2})=(1,~4)\ ,~\eta_2/\eta_1=0.25\ ,~\kappa=0\ , \lambda_{1,2}= 1 \ , e = {\frac{1}{\sqrt{2}}}\ ,
\end{align}
%
as reference values. Due to
$\eta_2/\eta_1 \ll 1$, $\phi_1$ obtains the VEV earlier than $\phi_2$ in the evolution.

We observed the formation of the compensated, the uncompensated, and the global strings after $U(1)$ symmetry breaking. 
Here, the global string corresponds to 
the strings
with $n_1 =0$, which appears when 
$\phi_2$ obtains the expectation value.
As the gauge boson has become massive by that time,
the global string does not have 
gauge flux at its core.
In our simulation, we found that the winding number of the formed global strings are only $\pm 1$.
Accordingly, the formed global strings have the domain-wall number $N_\mathrm{dw} = 1$.%
\footnote{For a model with $|q_1| \neq 1$ ($q_1 \in \mathbb{Z}$), the global string with $n_2 = 1$, has $N_\mathrm{dw} = \pm q_1$.}

The simulation also showed that most of the uncompensated strings evolve into the compensated strings by absorbing or releasing the global strings.
Such behavior is understood by the long-range forces between uncompensated and global strings, as discussed in \cite{Hiramatsu:2019tua}.
Thus, the 2+1\,D simulation
suggested that the string network evolves into the one with only the compensated strings ($N_{\mathrm{dw}} = 0$) 
and the global strings ($N_{\mathrm{dw}}=\pm 1$).

In our 2+1\,D simulation, we found that the strings with $|N_{\rm dw}|>1$ disappear at late times, which implies that the model might be free from the axion domain-wall problem.
However, this is not conclusive, because strings reconnect in the 3+1\,D spacetime, which cannot be 
simulated in the 2+1\,D simulation.
In the following, we perform the 3+1\,D simulation.

\section{Colliding two strings}
\label{sec:collision}
In this section, we simulate the collision between a compensated string and an uncompensated string with various winding numbers as a demonstration. We take the parameters%
\footnote{For this charge assignment, the $U(1)_\mathrm{global}$ symmetry does not appear as an accidental symmetry of the renormalizable theory. Still,
we assume the scalar potential
in Eq.\,\eqref{eq:potential} by assuming $U(1)_\mathrm{global}$ symmetry.}
%
\begin{align}
(q_1,~q_2)=(1,~2),~\eta_2/\eta_1=0.5,~\kappa=0,~\lambda_{1,2}=1,~e=\frac{1}{\sqrt{2}}\ .
\end{align}
%

\subsection{Simulation Setup}

The Euler-Lagrange equations of the scalar fields and the gauge field are given by, 
%
\begin{align}
&\mathscr{D}^\mu\mathscr{D}_\mu\phi_n=\frac{\partial V}{\partial\phi^{*}_n}\ ,\\
&\partial_\mu F^{\mu\nu}={ie}\sum_{n=1}^{2}q_n
\left[\phi^{*}_n
\mathscr{D}^\nu\phi_n-(\mathscr{D}^\nu\phi_n)^*\phi_n
\right]\ .
\end{align}
%
For the colliding string simulations, we take the Lorenz gauge, $\partial^\mu A_\mu=0$.

We prepare the initial configurations for the scalar fields and the gauge field, realizing the compensated and the uncompensated strings with $n_1=1$ and arbitrary $n_2$. Following our previous paper \cite{Hiramatsu:2019tua} (see its Sec.\,IIB), we first solve the one-dimensional boundary-value problem for $\phi_n$ and $A_\mu$, and obtain the axially symmetric static solution of the strings along the $z$-axis.
We put two of the resultant string solutions on the $x$-axis with a separation $d$ in the three-dimensional box, such as $(d/2,0,z)$ and $(-d/2,0,z)$. We rotate them by $\pm\alpha/2$ with respect to the $x$-axis so that the relative angle between them becomes $\alpha$. We also perform the Lorentz-boost for each string with velocity $(v_1,v_2)=(-v,+v)$ respectively so that they collide with each other at the coordinate origin of the box. 
Taking $d$ much larger than the width of the strings, i.e., $d \gg \order{\eta_{1,2}^{-1}}$,
we approximate the initial configuration with the two Lorentz-boosted strings by $\phi_n(t_0,\xx)=\phi^{(1)}_n(\xx)\phi^{(2)}_n(\xx)/\eta_n$ ($n=1,2$) with $A_\mu(t_0,\xx)=A^{(1)}_\mu(\xx)+A^{(2)}_\mu(\xx)$. Here $\phi^{(p)}_n(\xx)$ and $A^{(p)}_\mu(\xx)$ ($p=1,2$) represent each field configuration in a Lorentz-boosted string at  $(d/2,0,z)$ and at $(-d/2,0,z)$, respectively.
In the following simulations, we take $d=20\eta_1^{-1}$, $\alpha=0.2\pi$ and $v=0.7$, except for Fig.\,\ref{fig:string_bridge_q12} in which we set $\alpha=0.06\pi$.

In our simulations, we impose the boundary conditions on each side of the 3\,D spatial box so that the scalar and gauge fields on them are always given by the superposition of the Lorentz-boosted strings constructed above. Hence, we must terminate the simulation when any effects from the impact point of the collision come to the boundaries.

\subsection{Result}

\begin{figure}[t]
\begin{center}
 \begin{minipage}{.38\linewidth}
  \includegraphics[width=\linewidth]{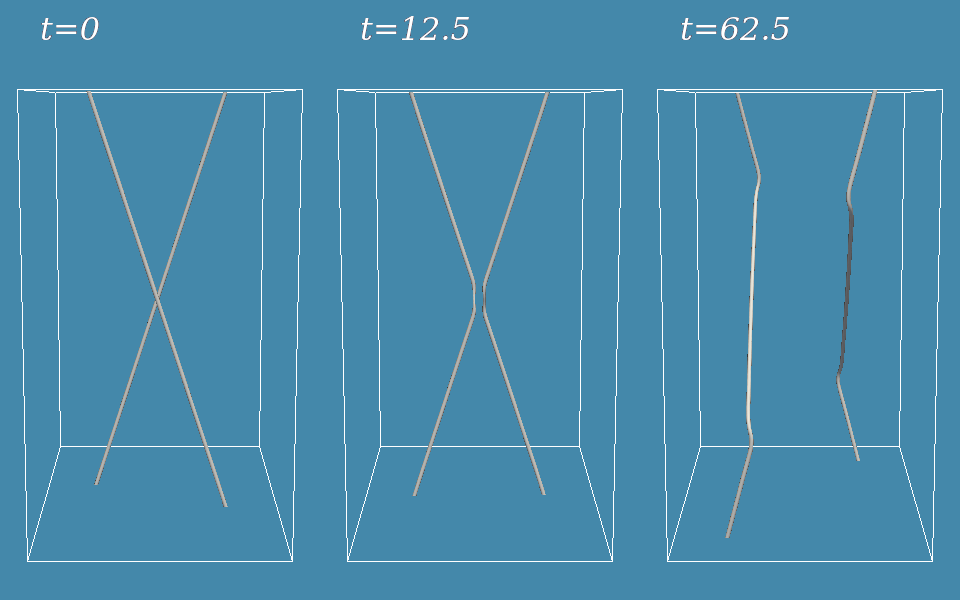}
 \end{minipage}
 \begin{minipage}{.57\linewidth}
  \includegraphics[width=\linewidth]{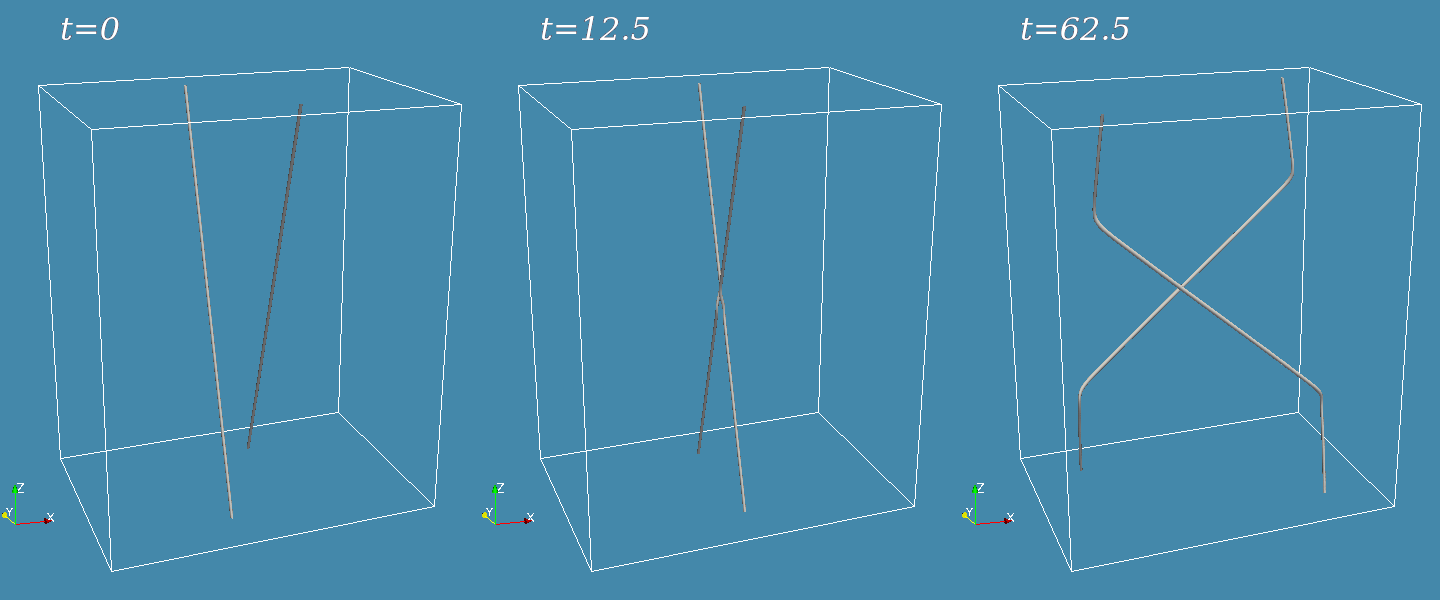}
 \end{minipage}
 \begin{minipage}{.38\linewidth}
  \includegraphics[width=\linewidth]{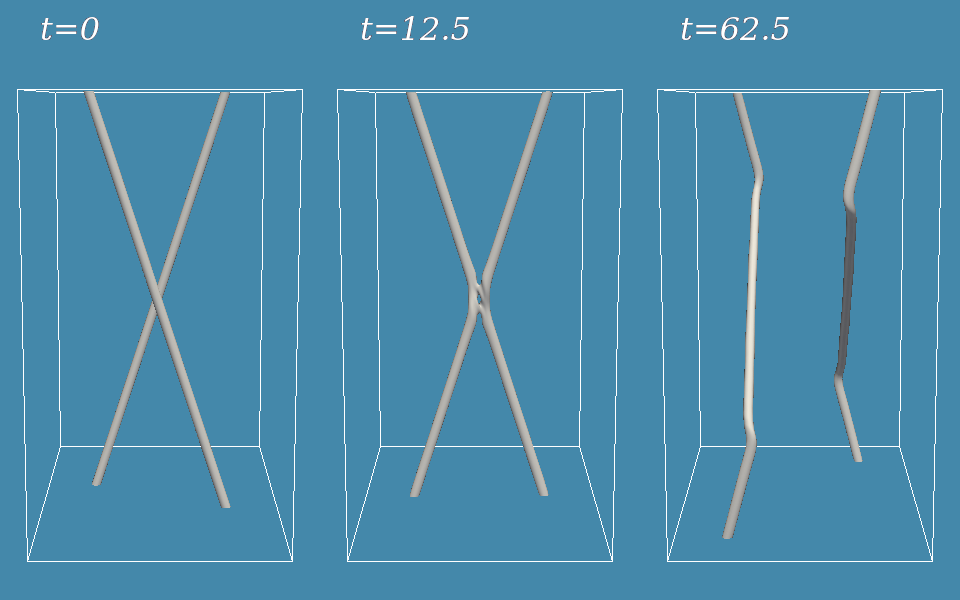}
 \end{minipage}
 \begin{minipage}{.57\linewidth}
  \includegraphics[width=\linewidth]{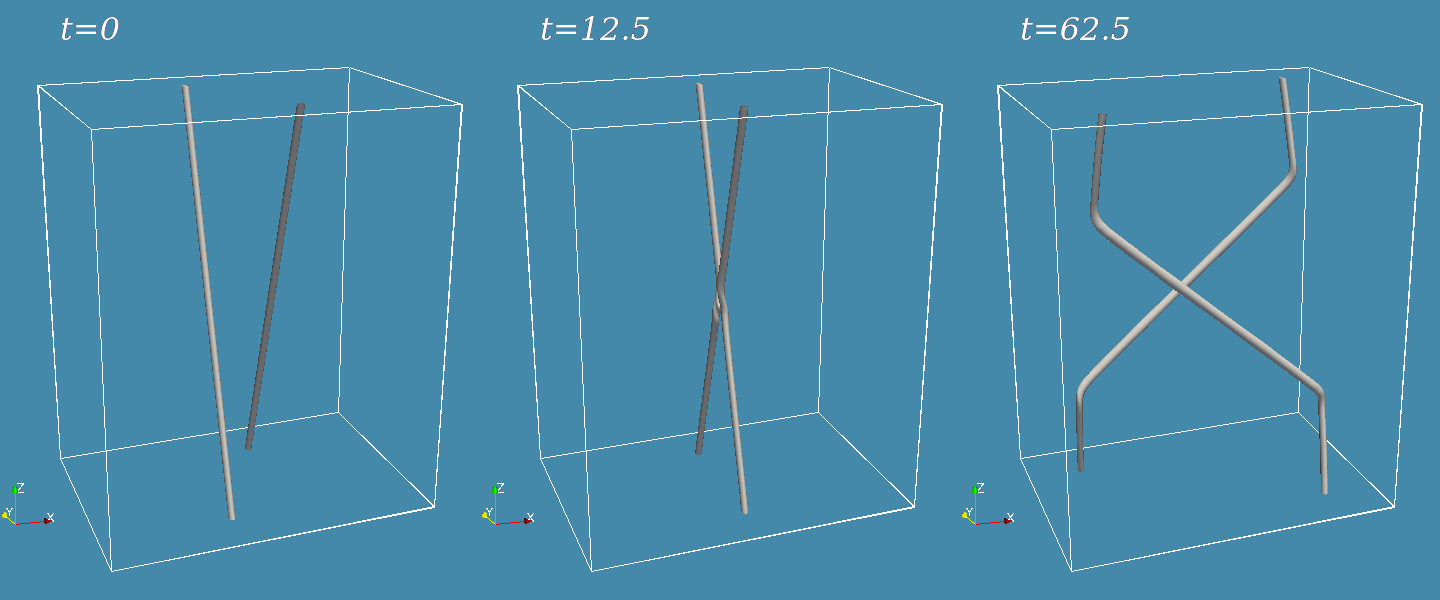}
 \end{minipage}
 \end{center} 
  \caption{
\sl \small \raggedright 
Collision between the compensated strings with $(n_1,n_2)=(1,2)$. The upper (lower) panels show the isosurface with $|\phi_1|=\eta_1/2$ ($|\phi_2|=\eta_2/2$). For each panel, we also show the same snapshot from a different view (left and right).}
\label{fig:colstr12} 
\end{figure}

\noindent
${\bf Case\,0:(n_1,n_2)=(1,2)}$\\
Before discussing the collision with the uncompensated string collision, let us see the collision between two compensated strings. The snapshots of the collision are shown in Fig.\,\ref{fig:colstr12}%
\footnote{Supplemental materials are available at {\tt http://numerus.sakura.ne.jp/research/open/NewString3D/}.}. The upper (lower) figures show the string  configurations made by $\phi_1$ ($\phi_2$). We observe that the two compensated strings simply reconnect and go away from each other as in the conventional Abelian-Higgs model.

\vspace{.5cm}

\begin{figure}[t]
\begin{center}
  \begin{minipage}{.48\linewidth}
  \includegraphics[width=\linewidth]{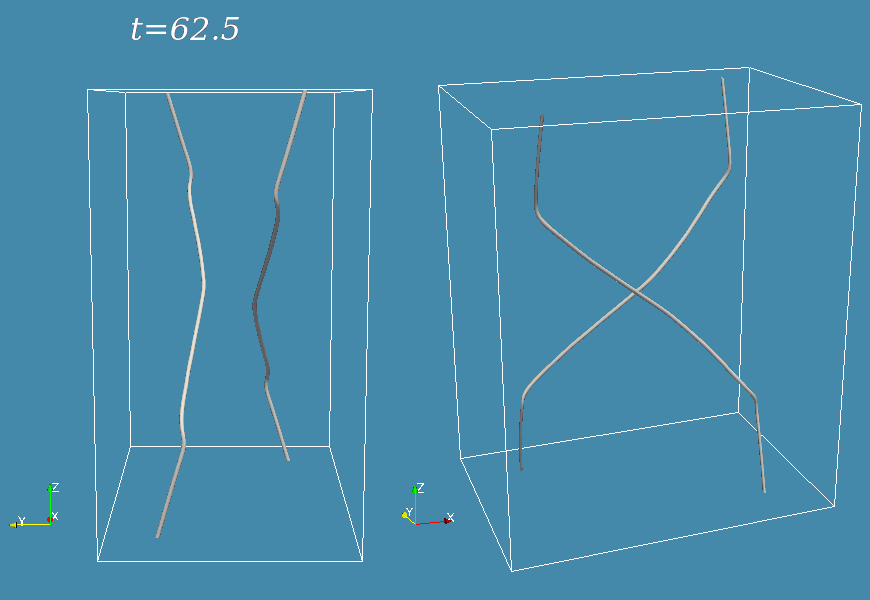}
 \end{minipage}
  \begin{minipage}{.48\linewidth}
  \includegraphics[width=\linewidth]{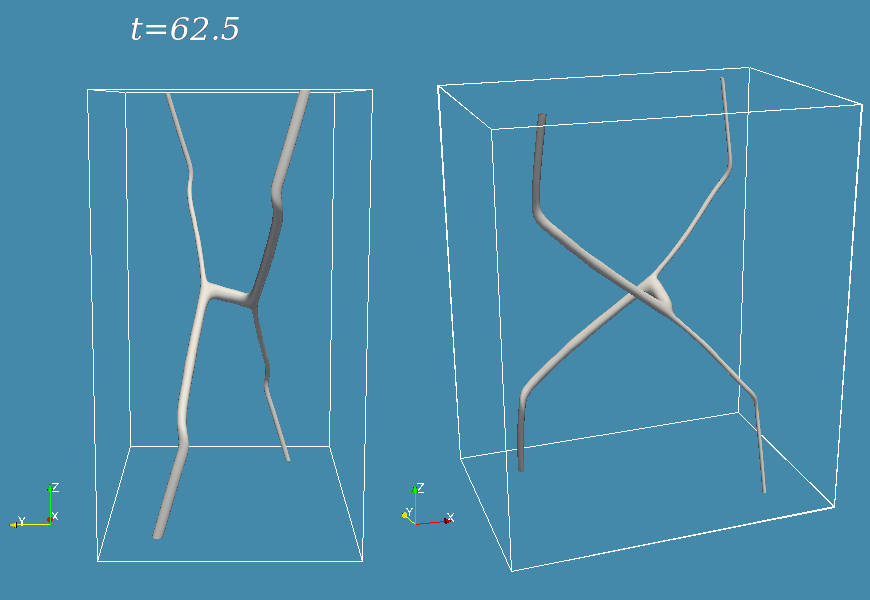}
 \end{minipage}
 \end{center} 
  \caption{
\sl \small \raggedright 
Collision between a compensated string with $(n_1 ,n_2)=(1,2)$ and an uncompensated string with $(n_1,n_2)=(1,1)$ at late times. The left (right) panels show the isosurface with $|\phi_1|=\eta_1/2$ ($|\phi_2|=\eta_2/2$).}
\label{fig:colstr11}
\end{figure}

\noindent
${\bf Case\,1: (n_1,n_2)=(1,1)}$\\
In Fig.\,\ref{fig:colstr11}, we show the snapshots of the collision between a compensated string and an uncompensated string with $(n_1,n_2)=(1,1)$ at the late time. 
As shown in the left panel, the $\phi_1$-strings%
\footnote{We call the string made by $\phi_1$ ($\phi_2$) the $\phi_1$-string ($\phi_2$-string).}
reconnect and move away as in the previous case. On the other hand, as shown in the right panel, the $\phi_2$-strings do not simply go away. 
After the collision, the $n_2 = 1$
$\phi_2$-string partially reconnects with $n_2=2$ $\phi_2$-string and form a combined system connected by a $n_2=1$ $\phi_2$-string. 

\begin{figure}[tbp]
\begin{center}
\begin{minipage}{.35\linewidth}
  \includegraphics[width=\linewidth]{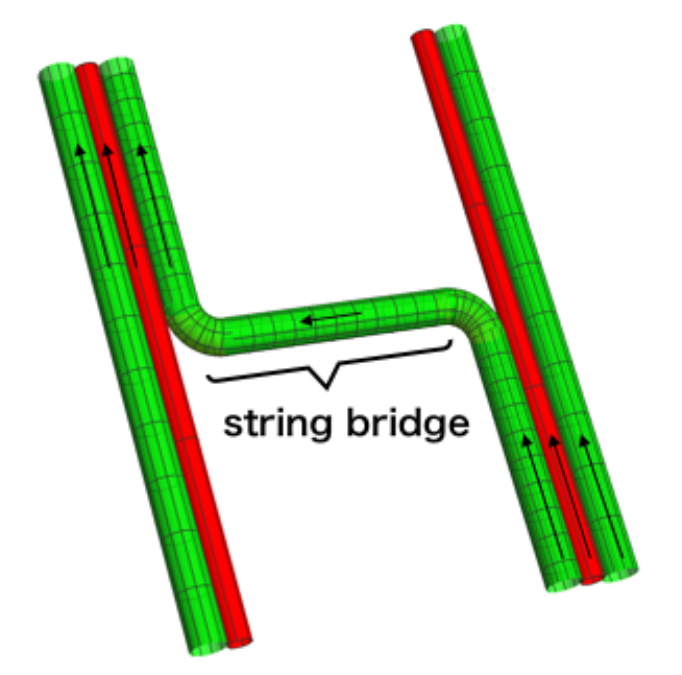}
 \end{minipage}
 \end{center} 
  \caption{
\sl \small \raggedright 
A schematic picture of the string system after the collision. The red rod is a $\phi_1$-string with $n_1=1$. The green rod is a $\phi_2$-string with $n_2=1$. The arrows on the strings denote the magnetic flux in the strings. We call the string segment connecting two strings the string bridge.
The string bridge is formed by the reconnection of the bunch of the strings. The strings are cut for drawing figures.}
\label{fig:string_bridge_q12}
\end{figure}

A schematic picture of this final state is shown in Fig.\,\ref{fig:string_bridge_q12}. 
One red rod and one green rod are the $\phi_1$-string with $n_1=1$ and the $\phi_2$-string with $n_2=1$, respectively. The direction of the arrow denotes the direction of the magnetic flux of the strings. We call the string segment connecting two strings the ``string bridge''. The string bridge is formed by the reconnection of the bunch of the strings with different $n_2$.

\begin{figure}[htbp]
\begin{center}
  \begin{minipage}{.40\linewidth}
  \includegraphics[width=\linewidth]{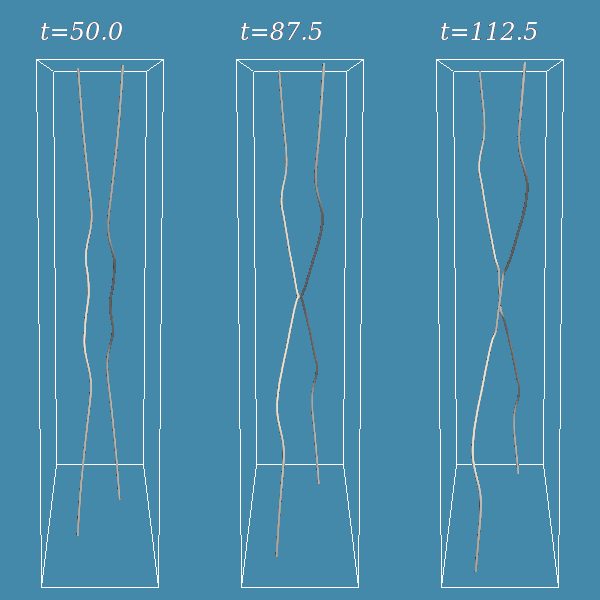}
 \end{minipage}
  \begin{minipage}{.40\linewidth}
  \includegraphics[width=\linewidth]{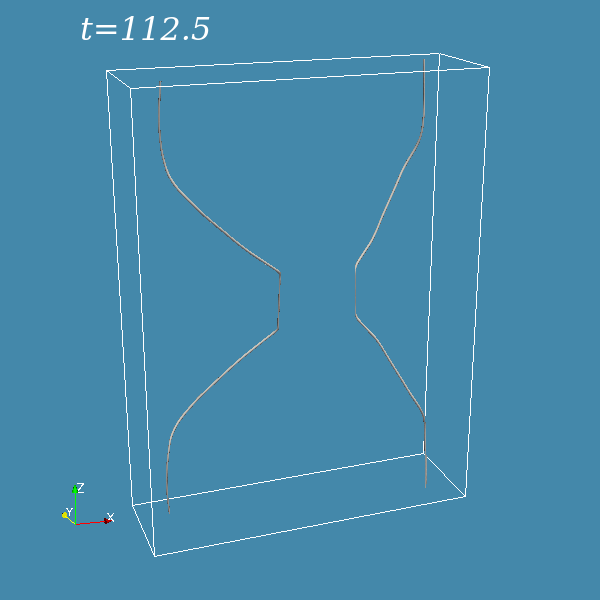}
 \end{minipage}
  \begin{minipage}{.40\linewidth}
  \includegraphics[width=\linewidth]{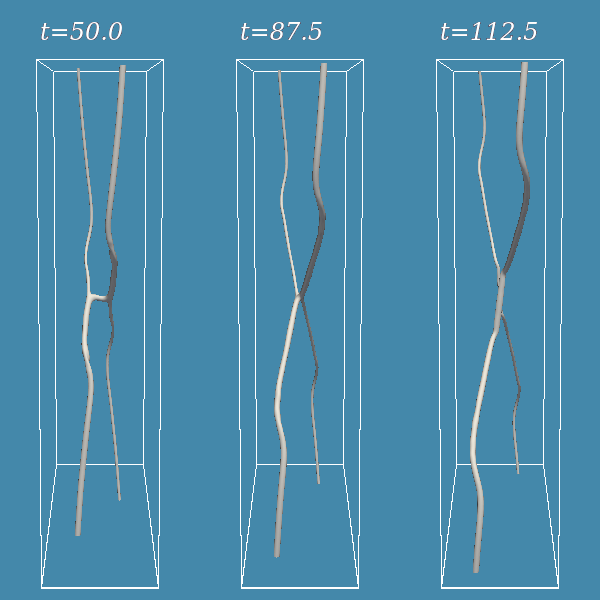}
 \end{minipage}
  \begin{minipage}{.40\linewidth}
  \includegraphics[width=\linewidth]{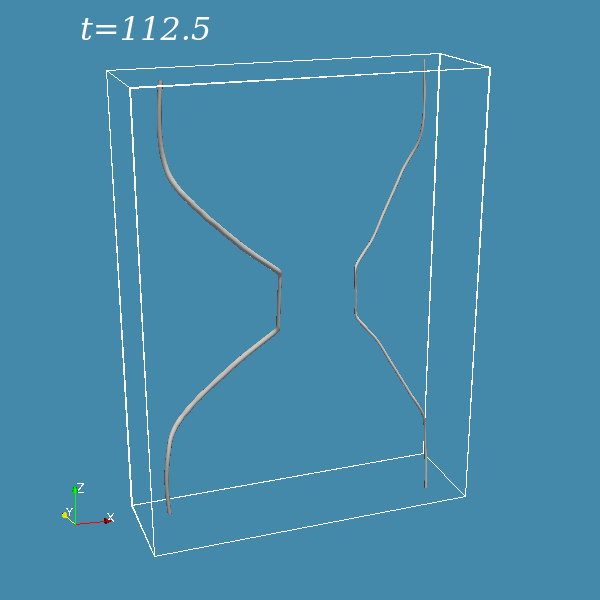}
 \end{minipage}
 \end{center} 
  \caption{
\sl \small \raggedright 
Collision between a compensated string with $(n_1 ,n_2)=(1,2)$ and an uncompensated string with $(n_1,n_2)=(1,1)$ for $\alpha=0.06\pi$. The notation is the same as Fig.\,\ref{fig:colstr12}. After the first reconnection before $t=50.0\eta_1^{-1}$, the second reconnection takes place at $t=87.5\eta_{1}^{-1}$.}
\label{fig:double_rec} 
\end{figure}

\begin{figure}[tbp]
\begin{center}
\begin{minipage}{\linewidth}
  \includegraphics[width=1.1\linewidth]{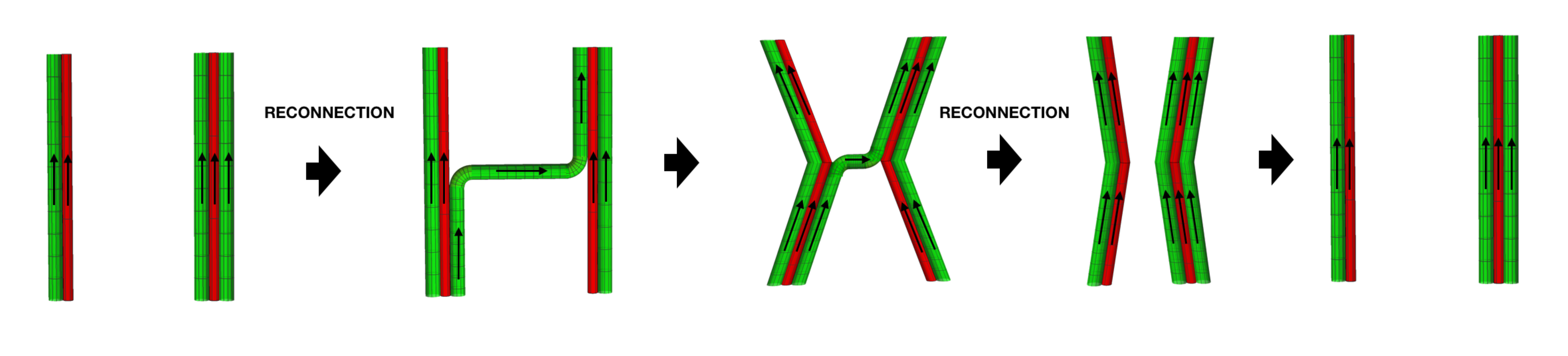}
 \end{minipage}
 \end{center} 
  \caption{
\sl \small \raggedright 
A schematic picture of the double reconnection of the strings for Fig.~\ref{fig:double_rec}. The red rod is a $\phi_1$-string with $n_1=1$. The green rod is a $\phi_2$-string with $n_2=1$. The arrows on the strings denote the magnetic flux in the strings. The strings are cut for drawing figures.}
\label{fig:1211_collision}
\end{figure}

When the collision angle, $\alpha$, is small,
we find that even the second reconnection takes place (Fig.\,\ref{fig:double_rec}). 
By the tension of the string bridge, two bunches of the strings do not leave away, and they are attracted to each other after the first reconnection. 
Once they collide and reconnect again ($t=87.5$ in the figure), they divide into two bunches of the strings, which are the same string bunches before the first collision.
Then, they eventually move away from each other as if they pass through each other.
See Fig.~\ref{fig:1211_collision} for the schematic diagram.

\begin{figure}[htbp]
\begin{center}
  \begin{minipage}{.48\linewidth}
  \includegraphics[width=\linewidth]{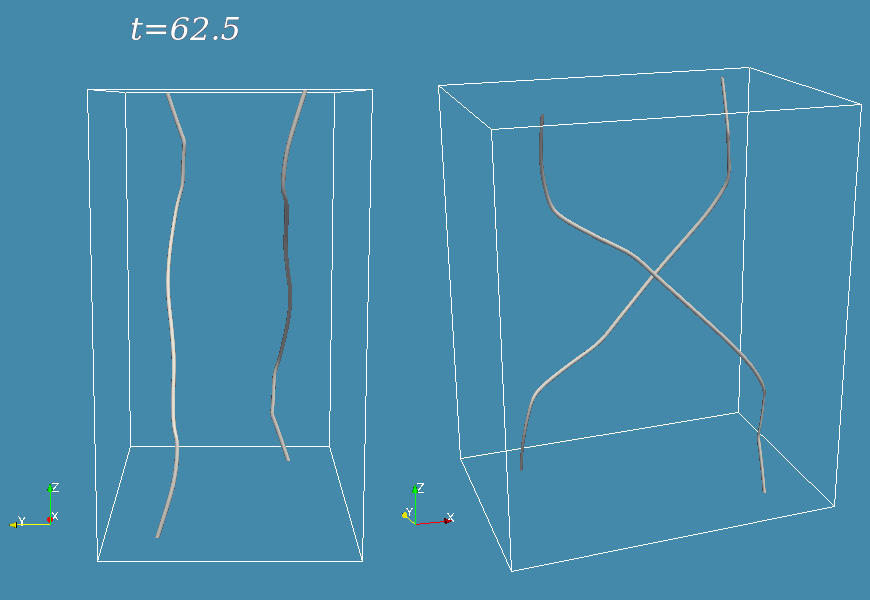}
 \end{minipage}
  \begin{minipage}{.48\linewidth}
  \includegraphics[width=\linewidth]{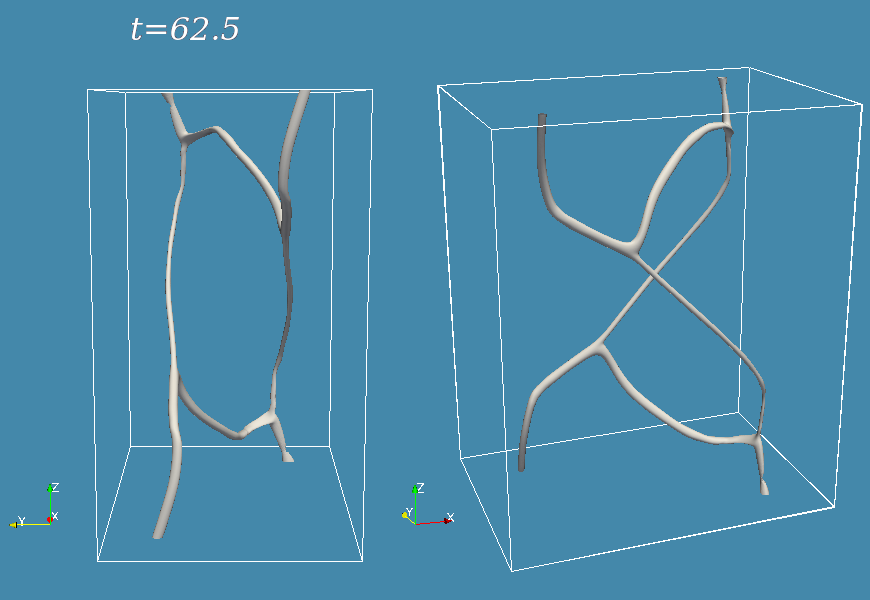}
 \end{minipage}
 \end{center} 
  \caption{
\sl \small \raggedright 
Collision between the compensated string with $(n_1 ,n_2)=(1,2)$ and the uncompensated string with $(n_1,n_2)=(1,0)$ at late times. The notation is the same as Fig.\,\ref{fig:colstr11}.}
\label{fig:colstr10} 
\end{figure}
\vspace{.5cm}

\begin{figure}[tbp]
\begin{center}
\begin{minipage}{\linewidth}
  \includegraphics[width=\linewidth]{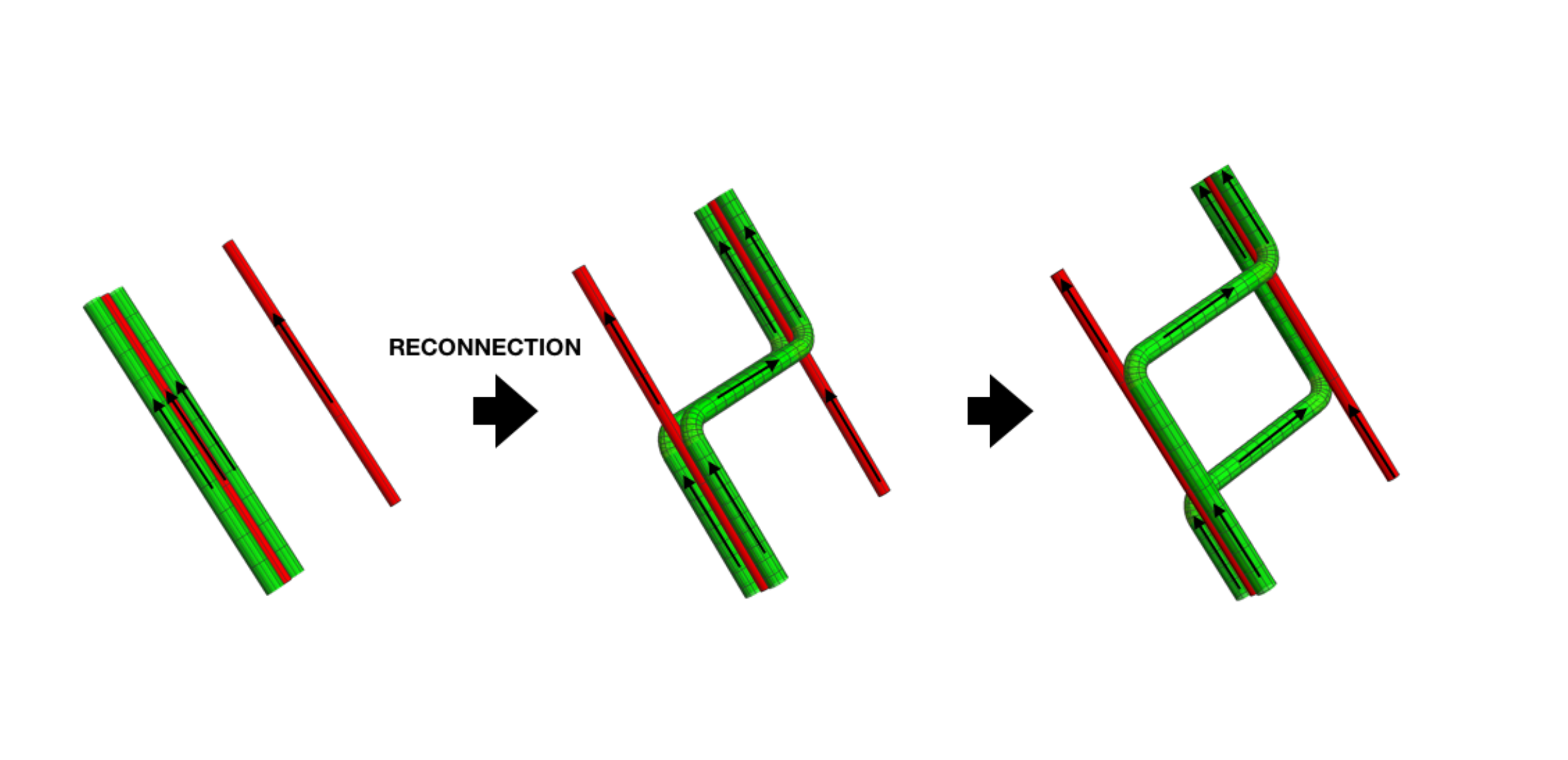}
 \end{minipage}
 \end{center} 
  \caption{
\sl \small \raggedright 
A schematic picture of collision between the compensated string with $(n_1 ,n_2)=(1,2)$ and the uncompensated string with $(n_1,n_2)=(1,0)$ for Fig.~\ref{fig:colstr10}. The red rod is a $\phi_1$-string with $n_1=1$. The green rod is a $\phi_2$-string with $n_2=1$. The arrows on the strings denote the magnetic flux in the strings. The strings are cut for drawing figures.}
\label{fig:1210_collision}
\end{figure}

\noindent
${\bf Case2: (n_1,n_2)=(1,0)}$\\
We also simulate the case with $(n_1,n_2)=(1,0)$ uncompensated string. As shown in Fig.\,\ref{fig:colstr10}, at the collision, the bunch of the $\phi_2$-strings reconnect and the string bridges consisting of two $\phi_2$-strings are formed.%
\footnote{Even for $n_2=0$, the radial component of $\phi_2$ takes non-trivial string configuration, i.e., $h_2(0) = 0$ and $h_2(\infty) = 1$, due to the non-trivial $A_\theta$ around the $\phi_1$-string. For details, see Ref.~\cite{Hiramatsu:2019tua}.}
After the formation of the bridges, the two bridges are separated and then keep a distance between them. This phenomenon is understood by the repulsive force between the two $\phi_2$-strings. See Fig.~\ref{fig:1210_collision} for the schematic diagram.

The colliding simulations have shown the formation of the string bridges, implying that 
the complicated string network emerges in the cosmological setup.
To see this and the fate of the uncompensated strings in the network at the late time, 
we perform the 3+1\,D classical lattice simulations in the radiation dominated universe in the next section.

\section{Cosmological evolution of string network}
\label{sec:simulation}
We explore the formation and the evolution of cosmic strings by the 3+1\,D classical lattice simulations. In this section, we take the parameters,
%
\begin{align}
\label{eq:parameters}
(q_{1},~q_{2})=(1,~4)\ ,~\eta_2/\eta_1=0.25\ ,~\kappa=0\ , \lambda_{1,2}= 1 \ , e = {\frac{1}{\sqrt{2}}}\ ,
\end{align}
%
as reference values. These parameters are the same as those in the previous work of the 2+1\,D simulation \cite{Hiramatsu:2019tua}
and also see Eq.\,\eqref{eq:parameters_d2}.

\subsection{Setup}
\begin{table}[t!]
\centering
\caption{
\sl \small  Simulation parameters. $H_{\rm in} (H_{\rm fin})$ are the Hubble parameter at the initial (final) time.}
\small{
\begin{tabular}{cc}
\hline
\hline
 \text{Grid size} &  $1024^3$\\
\text{Initial box size} &  $35H_{\rm in}^{-1}$ \\ 
\text{Final box size} &  $2H_{\rm fin}^{-1}$ \\
\text{Initial conformal time} & $2.01\eta^{-1}_1$ \\
\text{Final conformal time}& $35.1\eta^{-1}_1$\\
\text{Time step}& $2400$ \\ 
\hline
\hline
\end{tabular}
}
\label{tab:uni}
\end{table}%
We perform simulations in the radiation dominated universe whose spacetime metric is given as
%
\begin{align}
ds^2=a^2(\tau)(-d\tau^2+d\boldsymbol{x}^2)\ ,
\end{align}
%
where $\tau$ is the conformal time and $a(\tau)$ is the scale factor.
In the simulations, we use the temporal gauge, $A_0=0$, and hence, the field equations are given by
%
\begin{align}
\label{eq:eulers2}
&\ddot{\phi}_{n} + 2\mathcal{H}\dot{\phi}_{n} -\delta^{ij}\mathscr{D}_i\mathscr{D}_j\phi_n=-a^2V_{\phi^{*}_n}\ ,\\
\label{eq:eulerA2}
&\ddot{A}_k-\delta^{ij}\partial_iF_{jk}=-ia^2e\sum_n^2q_n
\left[\phi^{*}_n
\mathscr{D}_k\phi_n-(\mathscr{D}_k\phi_n)^*\phi_n
\right]\ .
\end{align}
%
Here, $\mathcal{H}=aH$ denotes the conformal Hubble parameter, $i,j,k=1,2,3$,  and the dot denotes the derivative with respect to the conformal time.

The initial conditions and the spatial periodic boundary conditions are the same as in our previous study \cite{Hiramatsu:2019tua}. 
As the initial conditions, we impose the random values for each grid point, $\phi(t_0,\xx)=\delta\phi(\xx)$, which obey the Planck distribution with the temperature $T=\sqrt{3}\eta_1$.
The time derivatives, $\dot{A}_{i}(t_0,\xx)$, are determined by solving the constraint equation, and as $A_{i}(t_0,\xx)$ can be freely chosen, we set $A_{i}(t_0,\xx)=0$. 
The time-evolution is calculated by the Leap-Frog method, and the spatial derivatives are approximated by the second-order finite-difference.
The simulation parameters are summarized in Table~\ref{tab:uni}.%

\subsection{Result}
\label{sec:result}
\begin{figure}[t]
\begin{center}
 \begin{minipage}{.48\linewidth}
  \includegraphics[width=\linewidth]{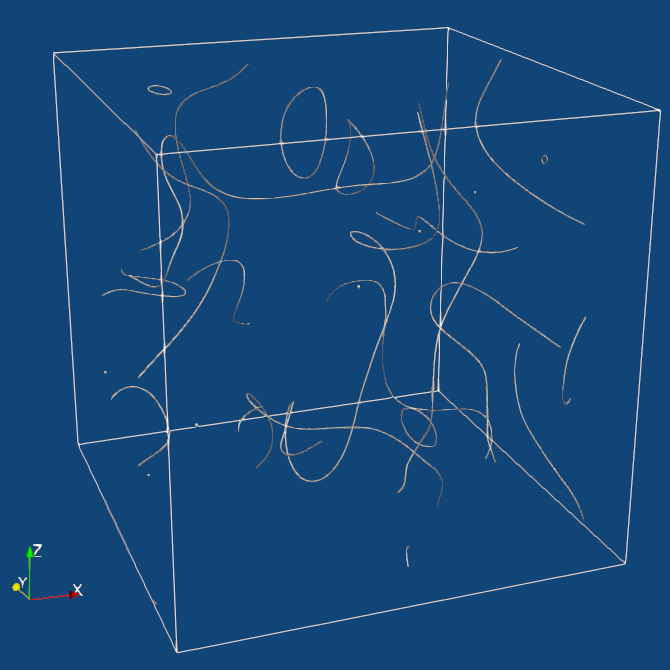}
 \end{minipage}
 \begin{minipage}{.48\linewidth}
  \includegraphics[width=\linewidth]{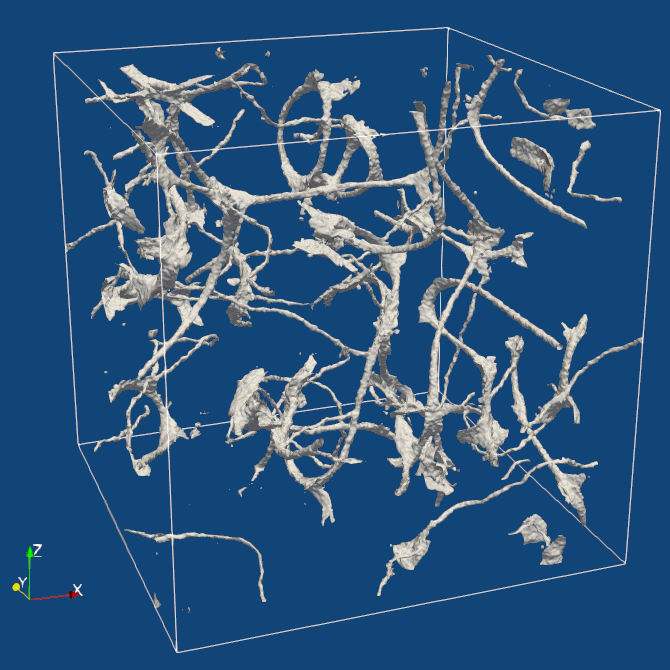}
 \end{minipage}
 \end{center} 
  \caption{
\sl \small \raggedright 
Snapshots of the isosurface with $|\phi_1|=4\eta_1/5$ ({\it left}) and $|\phi_2|=\eta_2/2$ ({\it right}) at the simulation end.}
\label{fig:snapshot} 
\end{figure}

We show the 3\,D spatial snapshots of the $\phi_1$- and the $\phi_2$-strings at the simulation end in Fig.\,\ref{fig:snapshot}.
The left (right) panel shows the configuration of the $\phi_{1(2)}$-strings.%
\footnote{See also {\tt http://numerus.sakura.ne.jp/research/open/NewString3D/} for supplemental material.}
As we assume $\eta_1=4\eta_2$, the width of $\phi_1$-string is roughly 1/4 of that of $\phi_2$-string and thus the $\phi_1$-strings look thinner. Note that the physical size of the lattice spacing reaches to the width of $\phi_1$-string at the simulation end, given as $\mathit{\Delta}x\sim 1.2\eta_1^{-1}$. 

We observe the larger number of strings in the $\phi_2$ sector than in the $\phi_1$ sector, indicating that there are many global strings with $n_1 = 0$ and $n_2 \ne 0$.
See, e.g., a $\phi_2$-string connecting from the bottom surface to the left one near the coordinate origin in the right panel; we cannot observe the accompanying string in the $\phi_1$ sector in the left panel.
This is again due to our arrangement of the size of the VEVs, $\eta_1 \gg \eta_2$.

\begin{figure}[tbp]
\begin{center}
\begin{minipage}{.48\linewidth}
  \includegraphics[trim={0.45in 0.5in 0.55in 0.5in},clip,width=\linewidth]{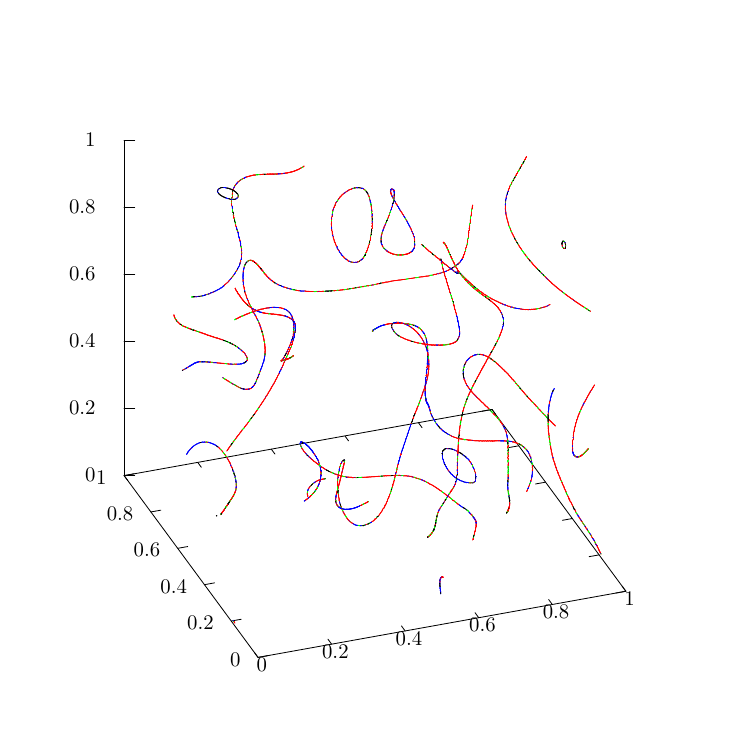}
 \end{minipage}
 \begin{minipage}{.48\linewidth}
  \includegraphics[trim={0.45in 0.5in 0.55in 0.5in},clip,width=\linewidth]{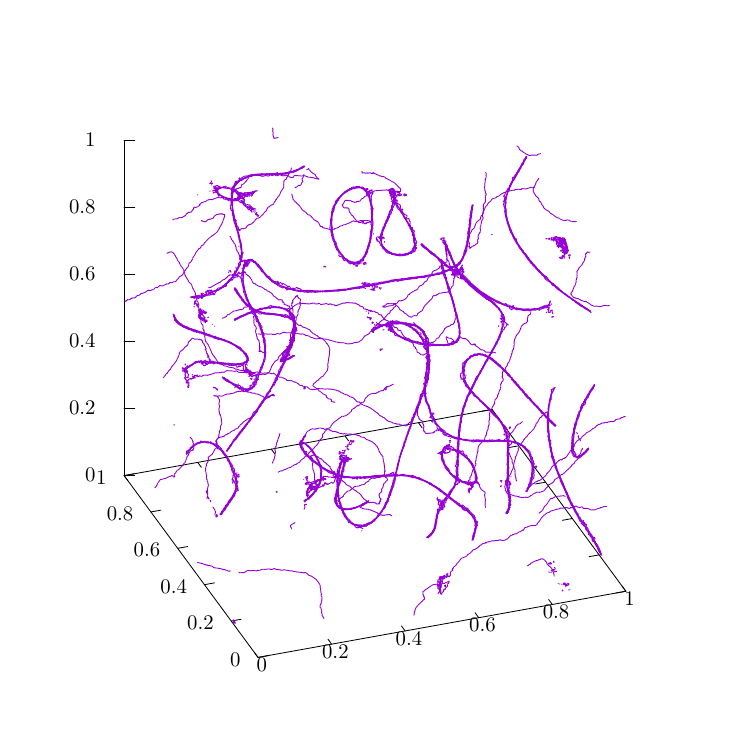}
 \end{minipage}
 \end{center} 
  \caption{
\sl \small \raggedright 
Fraction of strings with various $n_2$ for $(q_1,q_2;\eta_2/\eta_1)=(1,4;0.25)$ (\it left). 
String core of the $\phi_2$-string (\it right).   
The segments of the $\phi_1$-strings 
are colored in red for $(n_1,n_2)=(1,4)$, in blue for $(n_1,n_2)=(1,3)$, and in green for $(n_1,n_2)=(1,5)$.}
\label{fig:winding} 
\end{figure}

\begin{figure}[tbp]
\begin{center}
\begin{minipage}{0.7\linewidth}
  \includegraphics[width=\linewidth]{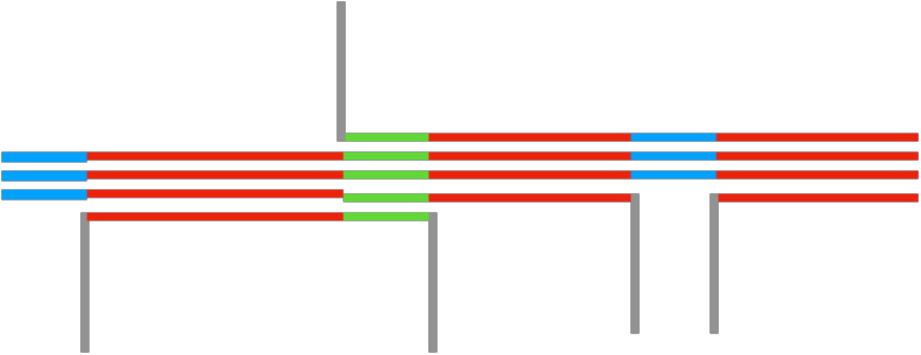}
 \end{minipage}
 \end{center} 
  \caption{
\sl \small \raggedright 
A schematic picture of typical fraction of string. Each line denotes the $\phi_2$-string. Red, blue, and green lines corresponds to $(n_1,n_2)=(1,4)$,~$(n_1,n_2)=(1,3)$, and $(n_1,n_2)=(1,5)$ strings, respectively. The gray lines show the string bridges. The strings are cut for drawing figures.}
\label{fig:string_fraction_schematic}
\end{figure}

In the right panel of Fig.\,\ref{fig:winding}, we show the string core of the $\phi_2$-string.
The higher winding strings with $n_2>1$ are represented as a bundle of strings with $n_2=1$, thus looking thick. The higher winding strings are expected to be accompanied with $\phi_1$-strings. To confirm this behavior, we count the number of $\phi_2$-strings along with a $\phi_1$-string, as shown in 
the left panel of Fig.\,\ref{fig:winding}.
Hereafter, $n_1$ denotes its absolute value. 
In the simulation, we find all the $\phi_1$-strings have $n_1=1$.
In this plot, a segment of a $\phi_1$-string coloured in red indicates that there are four $\phi_2$-strings along it distancing within $4\eta_1^{-1}$.%
\footnote{For the choice of the distance, see the Appendix~\ref{appsec:winding}.}
Along the $\phi_1$-string, we find that the sign of the winding numbers of the $\phi_2$-strings are the same with that of the $\phi_1$-string.
Thus, hereafter, we also take $n_2$ as its absolute value.
The segments colored in red are ``compensated'' since they satisfy $(n_1q_2-n_2q_1)=0$. 
On the other hand, there are three and five $\phi_2$-strings along the blue and green segments, respectively.  
The segments colored in blue or green are ``uncompensated'' with $n_1q_2-n_2q_1=\pm 1$,
respectively.
At the point where the color is changed, e.g. from red to blue, it is expected that there are either the string bridges as discussed in Sec.~\ref{sec:collision} or thin ambient global strings with no accompanying $\phi_1$-strings.
See Fig.~\ref{fig:string_fraction_schematic} for the schematic picture of an example string faction.

\begin{figure}[tbp]
\begin{center}
  \begin{minipage}{.4\linewidth}
  \includegraphics[width=\linewidth]{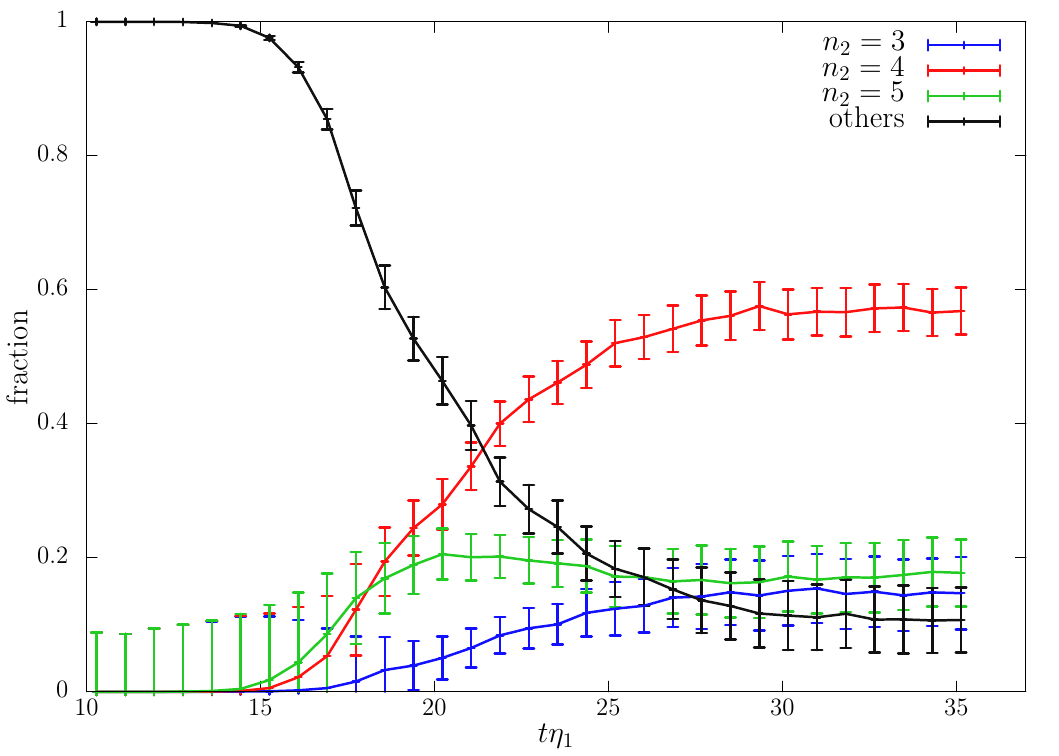}
 \end{minipage}
 \end{center} 
  \caption{
\sl \small \raggedright 
The time-evolution of the fractions of the strings, $R_{n_2}$, for $(q_1,q_2;\eta_2/\eta_1)=(1,4;0.25)$. 
}
\label{fig:fraction_0} 
\end{figure}

Let us see the time-evolution of the string segments of various kinds.
From the left panel of Fig.\,\ref{fig:winding}, we compute the lengths of strings with $n_1=1$ as $L_{\rm ng}$, 
which measures the total length of the non-global strings.
We also compute $L_{n_2}$, the length of the $\phi_1$-string segment which is accompanied 
by the $\phi_2$-strings with $n_2$.
In Fig.\,\ref{fig:fraction_0}, we show the time-evolution of the fractions,
\begin{align}
    \label{eq:Rn2}
    R_{n_2} = \frac{L_{n_2}}{L_{\rm ng}}\ .
\end{align}
The results are averaged over ten realizations and the error bars indicate the standard deviation.
In this plot, the red curve shows the time-evolution of $R_4$.
As it corresponds to the length of the compensated strings, we call it $R_c$ in the following. 
At late times, we find that about $60\%$ of the total non-global strings settle into the compensated strings.
This result is rather surprising in viewing the complicated reconnection processes in the 3+1\,D spacetime discussed in the previous section.
In the 2+1\,D simulation~\cite{Hiramatsu:2019tua}, the fraction $R_c$ settles to $R_c \sim 0.8$ at the end of the simulation.
The reduction of $R_c$ 
from $R_c\sim 0.8$ (in 2+1\,D) to $R_c\sim 0.6$ (in 3+1\,D) shows the effects of the complicated reconnection processes which make the combination process less efficient.

\begin{figure}[t]
\begin{center}
  \begin{minipage}{.4\linewidth}
  \includegraphics[width=\linewidth]{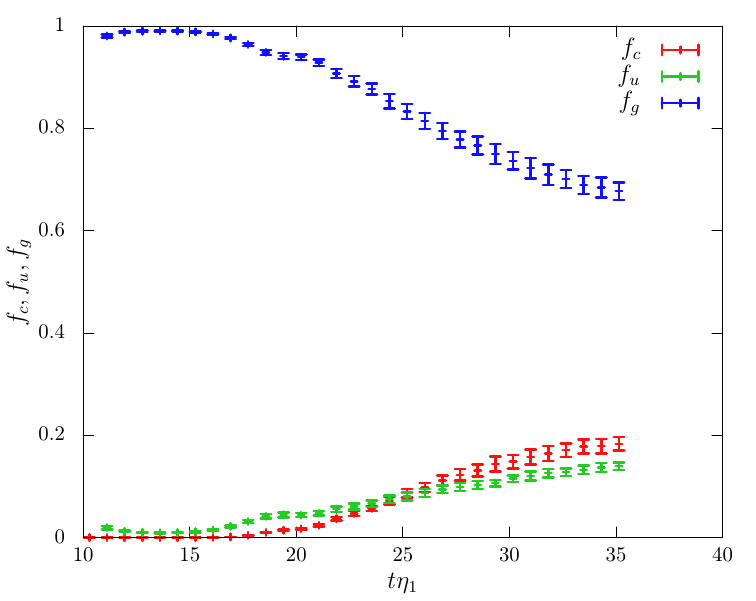}
 \end{minipage}
 \end{center} 
  \caption{\sl \small \raggedright 
The time-evolution of the fraction $f_c$, $f_u$, and $f_g$, for $(q_1,q_2;\eta_2/\eta_1)=(1,4;0.25)$.}
\label{fig:fraction} 
\end{figure}

In addition to the compensated/uncompensated strings, there are many $\phi_2$-strings which are not accompanied by the $\phi_1$-strings.
Those are the global strings, i.e., $n_1 = 0$.
In the simulations, we find no global strings with $n_2 > 1$.
To see the fraction of the global strings, we compute the length of the global string, $L_{\rm g}$, and define the fractions,
\begin{align}
    \label{eq:fcug}
    f_{\rm c} = \frac{L_{4}}{L_\mathrm{tot}}\ , 
    \quad
    f_{\rm u} = \frac{L_{n_2 \neq 4}}{L_\mathrm{tot}}\ , \quad
    f_{\rm g} = \frac{L_{\rm g}}{L_\mathrm{tot}}\ .
\end{align}
Here, the total string length is defined by
\begin{align}
    L_\mathrm{tot} = L_{\rm ng} + L_{\rm g} \ .
\end{align}
In Fig.\,\ref{fig:fraction}, we show the time-evolution of the fractions.
The figure shows that the string-network is dominated by the global strings
although the fraction decreases at late times. The latter fact indicates that 
some of the global strings are absorbed by the uncompensated strings to form the compensated strings, 
and some of them are reconnected with each other to form loops.%
\footnote{The reconnection and the loop formation of the global strings with $n_2 = 1$ 
take place in the usual way, regardless of the existence of $\phi_1$-strings. 
Hence, the global strings can disappear faster than the non-global strings.
}

\subsection{Physical Nambu-Goldstone Winding}

Let us discuss the time-evolution of the winding numbers 
of the gauge-invariant NG mode, $N_\mathrm{dw}$, defined in Eq.~(\ref{eq:def_Ndw}).
As we will discuss in the next section, 
the strings with $|N_\mathrm{dw}| > 1$ 
potentially cause the axion domain-wall problem if 
they remain abundantly at late times.
In the 2+1\,D simulation in~\cite{Hiramatsu:2019tua}, we found 
that the number of the 
strings with $|N_\mathrm{dw}|>1$ vanishes at late times. 
If this is the case, the axion model
based on the present set up can be 
free from the axion domain-wall problem (see discussion in Sec.\,\ref{sec:discussion}).
In the 3+1\,D, however, the string combination processes are less efficient, 
and hence, it is imperative to estimate how large fractions of the strings have $|N_\mathrm{dw}|>1$ at late times in the 3+1\,D simulation.

\begin{figure}[t]
\begin{center}
  \begin{minipage}{.4\linewidth}
  \includegraphics[width=\linewidth]{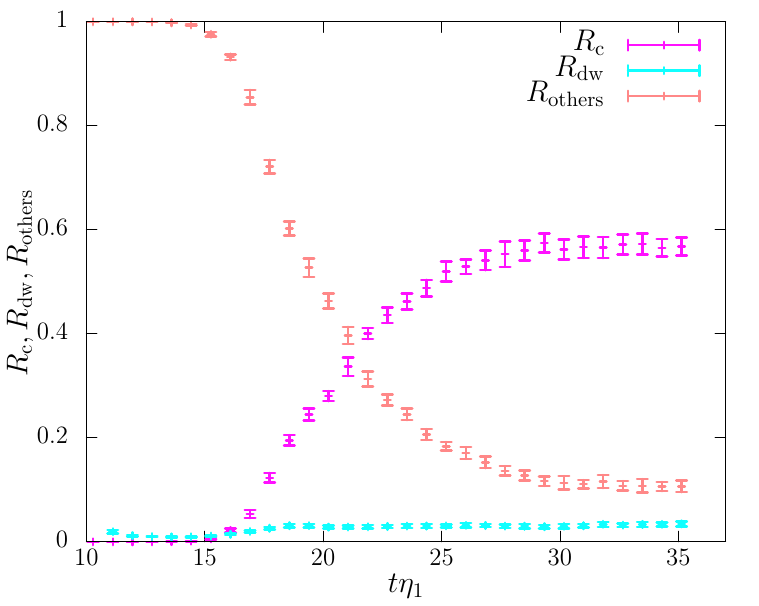}
 \end{minipage}
 \end{center} 
  \caption{\sl \small \raggedright 
The time-evolution of $R_c$, $R_\mathrm{dw}$, and $R_\mathrm{others}$, for $(q_1,q_2;\eta_2/\eta_1)=(1,4;0.25)$.}
\label{fig:fraction2} 
\end{figure}

To compare with the results of the 2+1\,D simulation, we compute the ratio,
\begin{align}
&R_{\rm dw}=\frac{L_{\rm others}}{L_{\rm tot}}\ ,\label{eq:def_Rdw}
\end{align}
in 3+1\,D, where $L_{\rm others}$ is the total length of the strings with $|N_\mathrm{dw}|>1$.%
\footnote{As mentioned earlier, no global strings with $n_2 > 1$ are formed, and hence, the global strings do not contribute to $L_\mathrm{others}$.}
This quantity is analogous to $R_\mathrm{dw}$ in~\cite{Hiramatsu:2019tua}.
In Fig.\,\ref{fig:fraction2}, 
we show the time-evolution of $R_\mathrm{dw}$ and $R_c$.
The figure shows that the $R_\mathrm{dw}$ is suppressed over the simulation time, but it does not vanish at
late times.
The non-vanishing $R_\mathrm{dw}$ is due to the complexity of the 
string network evolution in the 3+1\,D space-time. 

In the figure, we also show the evolution of the rescaled fraction,
\begin{align}
\label{eq:Rothers}
    R_\mathrm{others} = \frac{L_\mathrm{others}}{L_{\rm ng}} = \frac{L_\mathrm{tot}}{L_{\rm ng}} R_\mathrm{dw}\ ,
\end{align}
just for convenience.
As $L_\mathrm{tot}$ is dominated by the contribution of the global strings,
$R_\mathrm{others}$ is more suitable to see the fraction of the strings which have non-trivial NG winding around the non-global strings.

A caveat here is that $R_\mathrm{dw}$ and $R_\mathrm{others}$ do not 
count the uncompensated strings with $|N_\mathrm{dw}|>1$ precisely.
In the 2+1\,D simulation, the strings are point-like objects in the 2\,D space.
In this case, $R_\mathrm{dw}$ can be defined as the number fraction of the separate strings with $|N_\mathrm{dw}|>1$.
In the 3+1\,D simulation, on the other hand, the strings are connected as a network, and hence, the number of the $\phi_2$-strings, or the winding number $n_2$, changes along a $\phi_1$-string.
This is the reason why 
we define the various fractions based on the length of the segments in
Eqs.\,\eqref{eq:Rn2}, \eqref{eq:fcug}, \eqref{eq:def_Rdw}, and \eqref{eq:Rothers}.
Besides, the winding numbers of the segments are determined by counting the number of $\phi_2$-strings with $n_2=1$ along a $\phi_1$-string within $4\eta_1^{-1}$.
Thus, for example, 
$R_{n_2}$ is contaminated by 
the segments with $n_2'\neq n_2$, when the $\phi_2$-string around the $\phi_1$-string is loosely bounded or when there are some $\phi_2$-strings in the vicinity accidentally.
Therefore, $R_\mathrm{dw}$ should be taken as an upper limit of the fraction of the segment with $N_\mathrm{dw}>1$, since 
some portion of $R_{3,4,5}$ which contribute to $|N_{\mathrm{dw}}|\le 1$ are counted  as $R_\mathrm{dw}$.%
\footnote{As $R_{3}+R_{4}+R_{5} \gg R_\mathrm{dw}$, the relative uncertainty of
$R_\mathrm{dw}$ caused by the contamination is larger 
than that of $R_{3}+R_{4}+R_{5}$ (see also the Appendix \ref{appsec:winding}).}

\begin{figure}[tbp]
\begin{center}
  \begin{minipage}{.32\linewidth}
  \includegraphics[width=\linewidth]{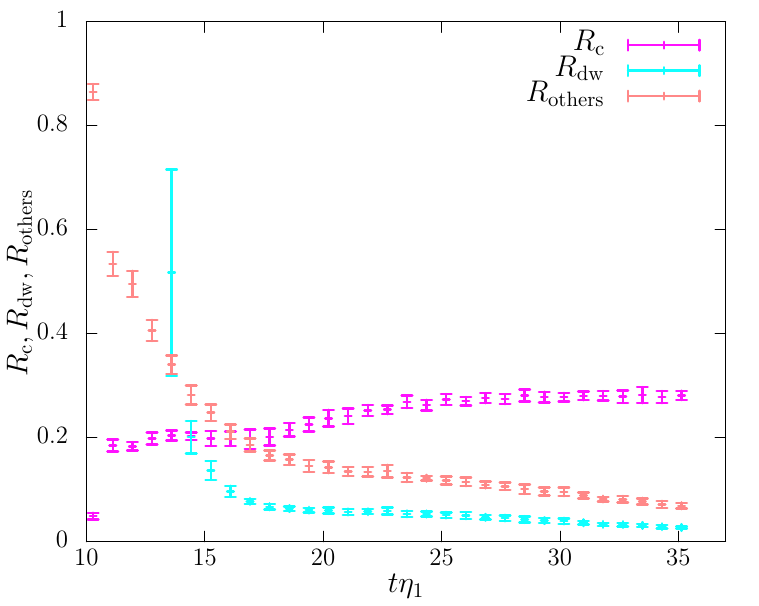}
 \end{minipage}
  \begin{minipage}{.32\linewidth}
  \includegraphics[width=\linewidth]{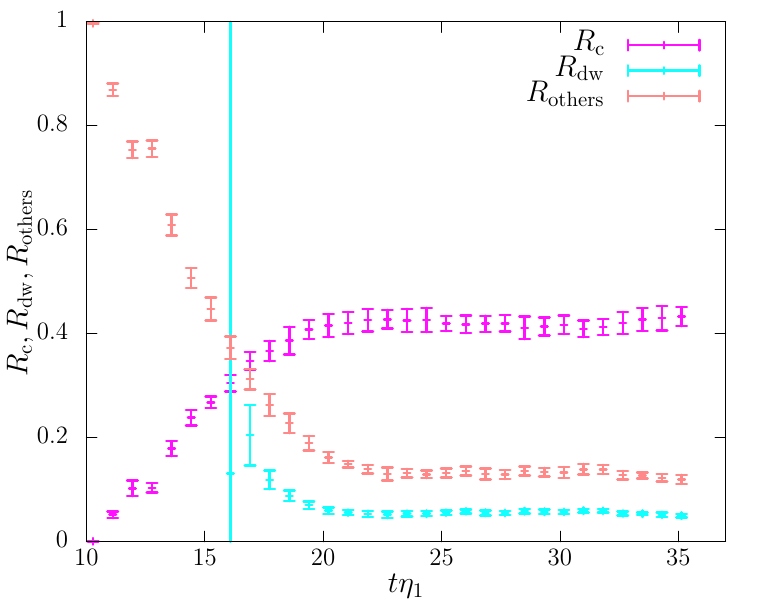}
 \end{minipage}
  \begin{minipage}{.32\linewidth}
  \includegraphics[width=\linewidth]{Figures/plot_d3m13c1-10_winding_ave_R_c4_0.pdf}
 \end{minipage}
 \end{center} 
  \caption{
\sl \small \raggedright 
The time-evolution of $R_{c,\mathrm{dw},\mathrm{others}}$
for various $\eta_2/\eta_1$.
We take $\eta_2/\eta_1=0.75,~0.50,~0.25$ from left to right. The other parameters are fixed to the values in Eq.\,\eqref{eq:parameters}. The rightmost panel is the same with Fig.\,\ref{fig:fraction2}.}
\label{fig:eta_dep} 
\end{figure}

Finally, let us show the parameter dependence.
In Fig.\,\ref{fig:eta_dep}, we show the time-evolution of $R_c,~R_{\rm dw}$ and $R_{\rm others}$ for $\eta_2/\eta_1=0.75,~0.50,~0.25$ (from left to right). 
We observe the increase of $R_{c}$ for a smaller $\eta_2/\eta_1$. This shows that more uncompensated strings are bounded into the compensated strings with $N_\mathrm{dw}= 0$ thanks to more abundant global strings at the phase transition of $\phi_2$ for $\eta_2\ll \eta_1$.
On the other hand, we do not observe significant tendencies
on $\eta_2/\eta_1$ for $R_{\rm dw}$ and $R_{\rm others}$ compared with the increase of  $R_c$.
These results indicate that $R_\mathrm{dw}$ and $R_{\rm others}$ could have sizable contributions from the contamination discussed above.

\begin{figure}[tbp]
\begin{center}
  \begin{minipage}{.32\linewidth}
  \includegraphics[width=\linewidth]{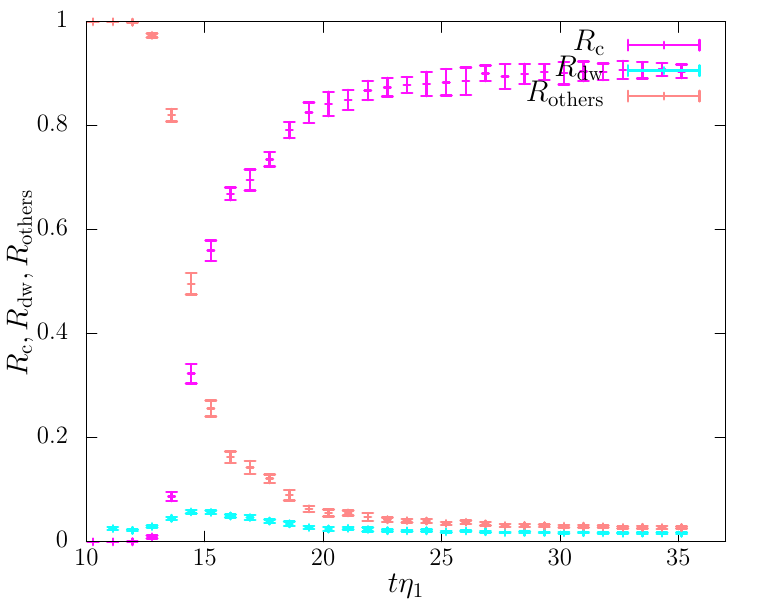}
 \end{minipage}
  \begin{minipage}{.32\linewidth}
  \includegraphics[width=\linewidth]{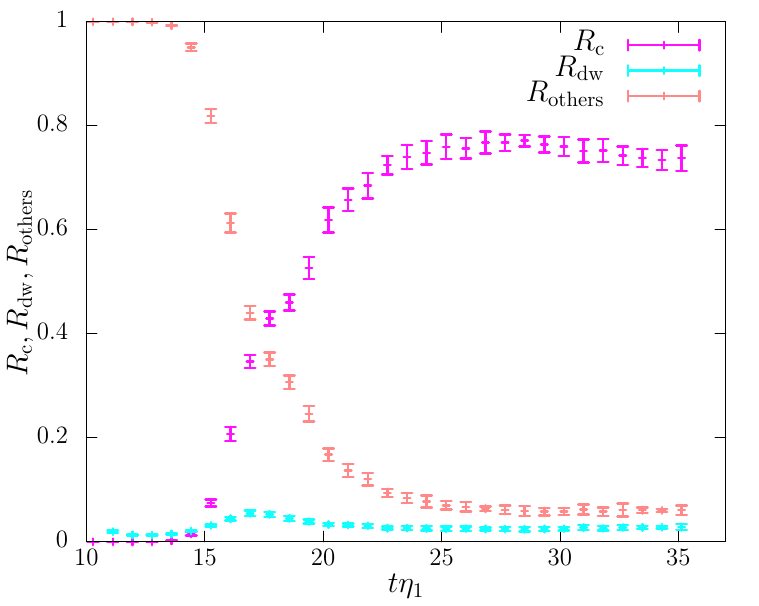}
 \end{minipage}
  \begin{minipage}{.32\linewidth}
  \includegraphics[width=\linewidth]{Figures/plot_d3m13c1-10_winding_ave_R_c4_0.pdf}
 \end{minipage}
 \end{center} 
  \caption{
\sl \small \raggedright 
The time-evolution of $R_{c,\mathrm{dw},\mathrm{others}}$
for various $q_2$.
We take $q_2=2,~3,~4$ from left to right. The rightmost panel is the same with Fig.\,\ref{fig:fraction2}.}
\label{fig:charge_dep} 
\end{figure}

We also show dependence on the charge ratio. Fig.\,\ref{fig:charge_dep} shows the time-evolution of $R_c,~R_{\rm dw}$ and $R_{\rm others}$ for $q_2=2,~3,~4$  (from left to right). We observe that $R_c$ decreases for a larger $q_2/q_1$. This is expected because 
the uncompensated strings require more $\phi_2$-strings to evolve into the compensated strings for a larger $q_2/q_1$.

Finally, let us comment that the case with the opposite charge ratio, $q_2/q_1=1/4$, 
while keeping the ratio of the VEVs, $\eta_2/\eta_1=0.25$.
In this case, we find that 
the strings with $(n_1,n_2)=(1,0)$ and $(0,1)$ remain at the end of the simulations,
although we do not show the results. 
Similar results are also obtained in the 2+1\,D simulation. This behavior can be understood that we need four $\phi_1$-strings and one $\phi_2$-string to form the compensated string.
This is difficult as $\phi_1$-strings are formed at $T\sim \eta_1$, while 
the $\phi_2$-strings are formed at much lower temperature, $T\sim \eta_2$.
As the global strings, formed at $T\sim \eta_2$, have $N_\mathrm{dw} = 4$, 
we expect $R_\mathrm{dw}$ is close to 1, for a model with $\eta_1 > \eta_2$
and $q_1 > q_2$.

\section{Implication to Axion domain-wall Problem}
\label{sec:discussion}
As we have seen in the previous section,
$R_\mathrm{dw}$ becomes small at late times  for $q_1=1$, $q_2>1$, and $\eta_1 \gg \eta_2$.
Besides, we found that the global strings with $n_2 = 1$ is the most abundant 
at late times.
In this section, we discuss implications 
of these results on the axion domain-wall problem.

First, let us consider the axion model
in which the Peccei-Quinn (PQ) symmetry is
identified to the $U(1)_\mathrm{global}$ symmetry~\cite{Fukuda:2017ylt,Fukuda:2018oco,Ibe:2018hir} (\cite{Kawasaki:2013ae} for review). 
As in the KSVZ axion model~\cite{Kim:1979if,Shifman:1979if}, we introduce $N_1$ and $N_2$ of the QCD colored vector-like multiplets, $(Q_{if_i},\bar{Q}_{if_i})$ ($i=1,2$, $f_i=1\mbox{--}N_i$), and couple them
to $\phi_1$ and $\phi_2^*$ via,
\begin{align}
   {\cal L}= \phi_1 \bar{Q}_1 {Q}_1  + \phi_2^*\bar{Q}_2 {Q}_2 + h.c.
\end{align}
Here, we suppress the Yukawa coupling constants and the flavor indices. 
The anomaly-free condition of the $U(1)_\mathrm{local}$ gauge symmetry requires,
\begin{align}
    N_1 q_1 - N_2q_2 = 0\ ,
\end{align}
which can be solved by%
\footnote{We define $q_1$ and $q_2$ being relatively prime numbers.}
\begin{align}
    N_m = \frac{N_1}{q_2} = \frac{N_2}{q_1}\ , \quad N_m \in \mathbb{Z}^{>0}\ .
\end{align}
The integer $N_m$ is a free parameter of the axion model.

Once the $U(1)$ symmetries are broken spontaneously, the vector-multiplets become massive, and they leave the coupling of the 
gauge-invariant NG boson in Eq.\,\eqref{eq:decomp} with the QCD gluon,
\begin{align}
{\cal L} = \frac{g_s^2}{32\pi^2}
\frac{N_m a}{F_a}
G\tilde{G}\ .
\label{eq:anomalous}
\end{align}
This shows that the gauge-invariant NG boson $a$ plays the role of the QCD axion.

As we have discussed in the previous section,
the string network for $q_1=1$, $q_2>1$, and $\eta_1 \gg \eta_2$ 
is dominated by the global strings 
(with only $n_2=1$).
The network also has a small contribution from the uncompensated segments with $|N_\mathrm{dw}|>1$.
Around these strings, the axion (i.e. the gauge-invariant NG boson) winds $N_\mathrm{dw}$ times in its domain defined in Eq.\,\eqref{eq:domain}.

When the temperature of the universe decreases below the QCD scale, the axion
obtains a periodic potential by the effect of Eq.\,\eqref{eq:anomalous}.
The period of the potential is $F_a/N_m$, and hence, it has a discrete $\mathbb{Z}_{N_m}$ shift symmetry.
Due to the axion winding around the strings,
the axion potential causes the energy contrasts   around the strings.
By remembering the periodicity of the axion potential, the energy contrasts have
the periodicity of $N_m \times |N_{\rm dw}|$ which results in $N_m \times |N_{\rm dw}|$ domain-walls attached to the strings.

For $N_m > 1$, the string-wall network is stable since the domain-walls formed in the above process are 
associated with 
the spontaneous breaking of the $\mathbb{Z}_{N_m}$ symmetry.
In this case, all the strings are attached by at least $N_m$ of the domain walls, and they are pulled into multiple directions.
Once stable domain-walls are formed, they immediately dominate the energy density of the universe and lead to the overclosure.
To avoid such a problem, we consider a model with $N_m = 1$
in the following.

For $N_m=1$, most strings are attached by 
only one domain-wall, since the string network is dominated by the global strings which have $|N_\mathrm{dw}|= 1$ in the present setup.
Since a string which are attached by only one domain-wall is pulled into one direction, it can annihilate with another string with opposite winding which is attached to the other side of the domain-wall~\cite{Vilenkin:1982ks,Hiramatsu:2012gg}.
Thus, the string network with $|N_\mathrm{dw}|=1$ does not cause the domain-wall problem.

Even for $N_m = 1$, however, the string network of the present model potentially results in  stable domain-walls since it has segments with $|N_\mathrm{dw}|>1$.
In fact, if the multiple domain-walls 
connect string segments with $|N_\mathrm{dw}|>1$, the strings are 
pulled into multiple directions 
and form a stable string-wall network.
As we have seen,
however, $R_\mathrm{dw}\ll 1$ and the most strings are the global string with $N_\mathrm{dw} = 1$ for $\eta_2/\eta_1 \ll 1$,
$q_1 = 1$, and $q_2 > 1$.
Thus, the stable string-wall networks 
made by the string segments with $|N_\mathrm{dw}|>1$ are surrounded by many global strings with $|N_\mathrm{dw}|=1$.
Therefore, the stable string-wall networks are immediately shredded when the global strings pass through the domain-wall, 
and disappear before they dominate the 
energy density of the universe.%
\footnote{The typical distance between the 
strings is of $\order{H^{-1}}$ where $H$ is the Hubble parameter at around the QCD transition. Thus, the time-scale of the dynamics of the string-wall network is of $\order{H^{-1}}$.}
These arguments strongly suggest that the axion model based on the model 
with $\eta_2/\eta_1\ll 1$,  $q_1=1$, $q_2 > 1$, and $N_m=1$ is free from the 
axion domain-wall problem.

Let us finally comment that the absence of the domain-wall problem cannot simply
result from the fact that the model has no discrete symmetry for $N_m = 1$.
For example, if we take $q_1/q_2 = 1/4$, we expect $R_\mathrm{dw}$ is close to 1, and the string network is dominated by the global strings with $N_\mathrm{dw} = 4$.
Thus, in this case, the model suffers from the domain-wall problem even in the absence of the discrete symmetry.

\section{Summary}
\label{sec:summary}
In this paper, we studied the formation and evolution of the string network in the model with one local $U(1)$ and one global $U(1)$ symmetries by performing the classical lattice simulations in the 3+1\,D spacetime. As shown in Sec.~\ref{sec:collision}, 
the reconnection processes of the present model are highly complicated.
In particular, the uncompensated strings are not simply reconnected, but yielding bridges which connect the 
two bunch of the strings (Fig.\,\ref{fig:string_bridge_q12}).
We also studied the time-evolution of the string network in the 3+1\,D spacetime.
Despite its complexity, we observed that more than $50\%$ of the uncompensated strings settle into the compensated strings even in
3+1\,D (Fig.\,\ref{fig:fraction_0} and Fig.\,\ref{fig:fraction}).

As a result of the network simulations, we found that the 
string-network is dominated by the global string (Fig.\,\ref{fig:fraction}) at late times.
Besides, we also found that the fraction of the string segments with $|N_\mathrm{dw}|>1$ is suppressed, i.e. $R_\mathrm{dw} \ll 1$ for $\eta_2/\eta_1 \ll 1$, $q_1=1$ and $q_2 > 1$ (Figs.~\ref{fig:fraction2} and \ref{fig:charge_dep}).
These results strongly suggest 
that the PQ axion model based on the present setup can be free from the axion domain-wall problem
when we take $\eta_2/\eta_1 \ll 1$ with  $q_1 = 1$, $q_2>1$, and $N_m = 1$.%
\footnote{Here, we assume that $U(1)$ spontaneous breaking takes place after inflation. If it happens before inflation, the string-network has been diluted and causes no axion domain-wall problem regardless of how the axion winds around the strings. Such a possibility causes the isocurvature problem, which puts upper limit on  
the Hubble parameter during inflation (see \cite{Kawasaki:2013ae} for review).}
This type of the axion model naturally explains the origin of the PQ symmetry from the gauge symmetry~\cite{Lazarides:1982tw,Barr:1982uj,Choi:1985iv,Fukuda:2017ylt,Fukuda:2018oco,Ibe:2018hir}.
The absence of the domain-wall problem 
makes this type of the axion more attractive.

There are several open questions. For the evolution of the string network, it is important to see whether the network follows the scaling law or not. Due to the formation of the string bridges, the scaling law may not be observed even for the time duration simulated in this paper. Another question is the production of the gravitational waves.
Their spectrum may have a feature arising from the complexity of the network associated with the combination process of the uncompensated strings, and therefore there is a possibility that we can distinguish them from the gravitational waves from
the other kind of strings like Abelian-Higgs strings.
In the present work, we do not discuss these issues further and leave them for future studies.

\vspace{-.4cm}  
\begin{acknowledgments}
\vspace{-.3cm}
This work is supported in part by JSPS KAKENHI Grant Nos. 17H02878, and 18H05542 (M.I.);
World Premier International Research Center Initiative (WPI Initiative), MEXT, Japan (M.I.). 
M. S. thanks Kavli IPMU for their hospitality during the corona virus pandemic.
\end{acknowledgments}

\appendix

\section{Computation of winding number}
\label{appsec:winding}

\begin{figure}[htbp]
\begin{center}
 \begin{minipage}{.4\linewidth}
  \includegraphics[width=\linewidth]{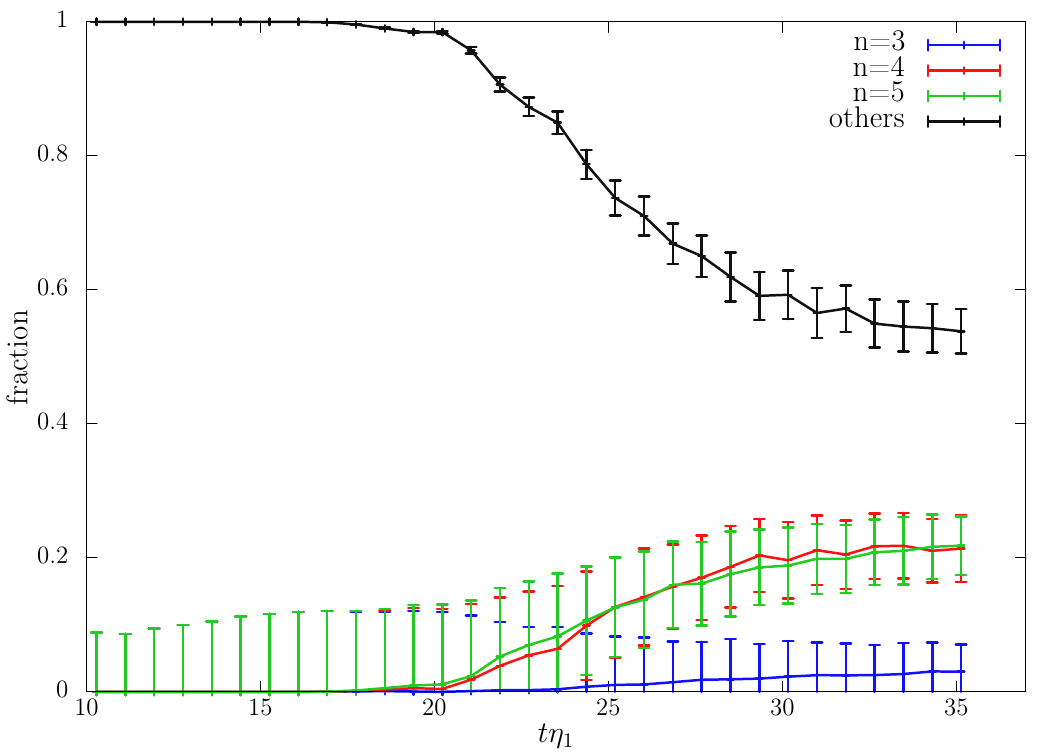}
 \end{minipage}
 \begin{minipage}{.4\linewidth}
  \includegraphics[width=\linewidth]{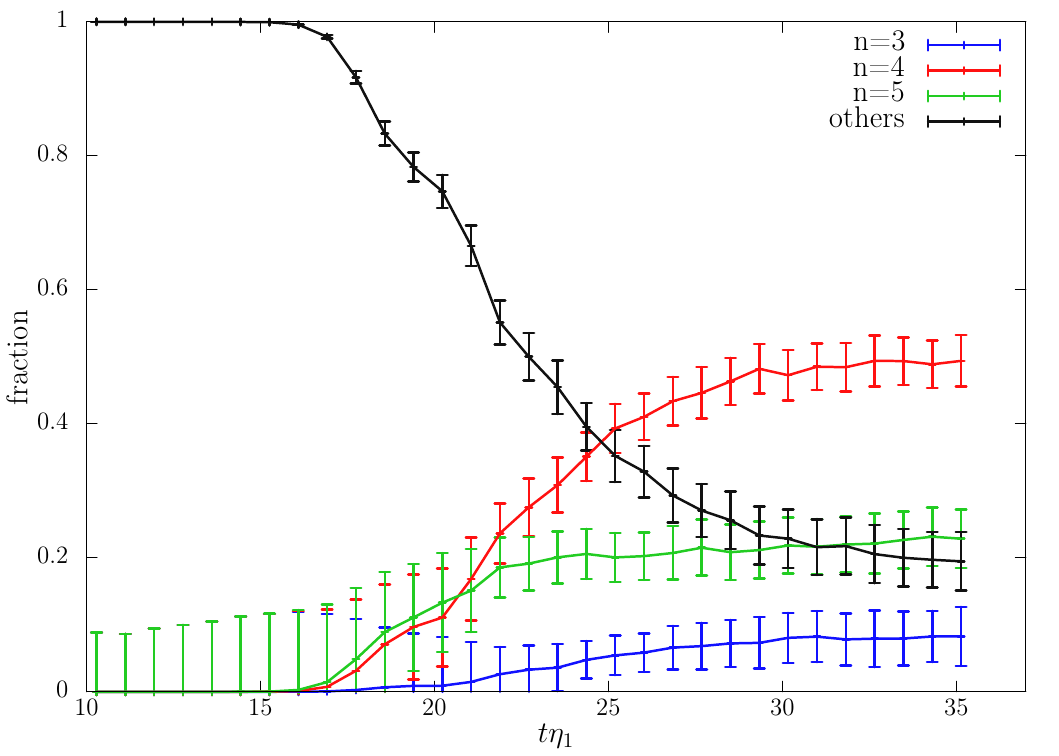}
 \end{minipage}
\begin{minipage}{.4\linewidth}
  \includegraphics[width=\linewidth]{Figures/plot_d3m13c1-10_winding_ave_c4_0.pdf}
 \end{minipage}
 \begin{minipage}{.4\linewidth}
  \includegraphics[width=\linewidth]{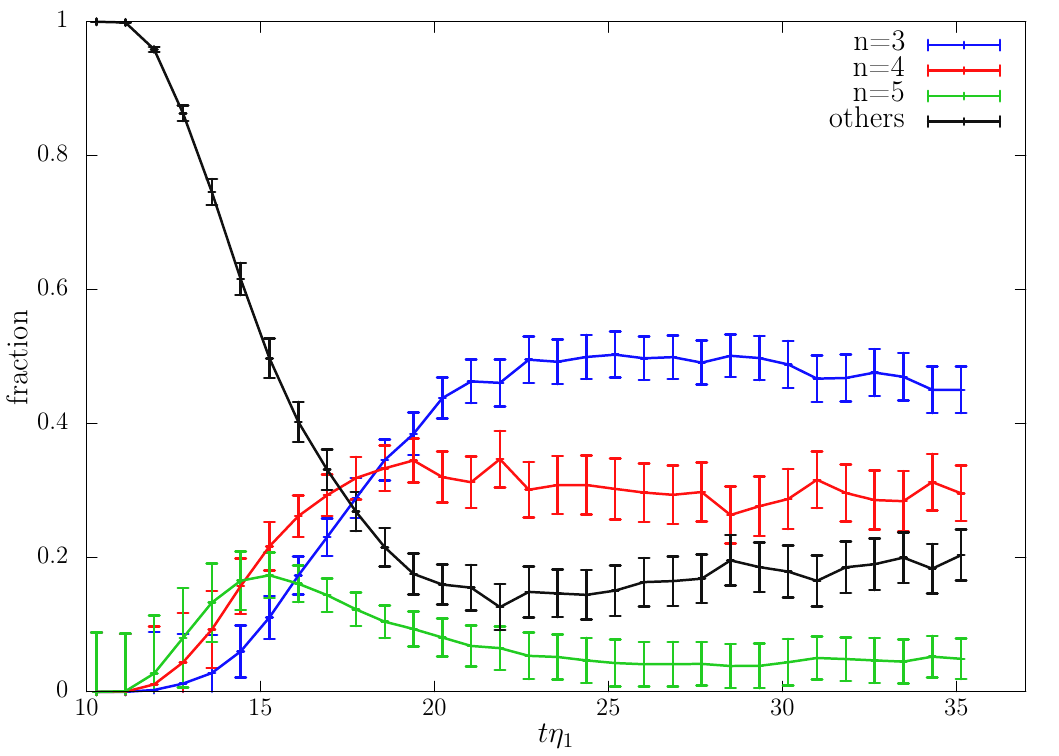}
 \end{minipage}
 \end{center} 
  \caption{
\sl \small \raggedright 
The time-evolution of $R_{3,4,5}$
and $R_\mathrm{others}$ for various choices of $s$. The top-left, top-right, bottom-left, and bottom-right figures show the cases with $s=16\eta_{1}^{-1}$, $6\eta_{1}^{-1}$, $4\eta_{1}^{-1}$, and $2\eta_{1}^{-1}$, respectively. 
}
\label{fig:capture_radius}
\end{figure}

In Sec.~\ref{sec:simulation}, we computed the winding number of $\phi_{2}$-string along a $\phi_{1}$-string as shown in Fig.\,\ref{fig:winding}. 
The winding number is given as the number of $\phi_{2}$-strings with $n_{2}=1$ lying along a $\phi_{1}$-string within a separation $s$. In the main text, we fixed $s=4\eta_{1}^{-1}$. In Fig.\,\ref{fig:capture_radius}, we show the dependence of the computed winding numbers on the parameter $s$. The bottom-left panel in this figure shows the case with $s=4\eta_{1}^{-1}$ and is the same as Fig.\,\ref{fig:fraction_0}. 

The top-left, top-right, and bottom-right figures show the cases with $s=16\eta_{1}^{-1}$, $6\eta_{1}^{-1}$, and $2\eta_{1}^{-1}$, respectively. 
By increasing $s$ from $4\eta_1^{-1}$ to $16\eta_1^{-1}$, we find that 
$R_\mathrm{others}$ increases and $R_{3,4,5}$ decrease.
This trend can be explained by the overcounting of global strings around the $\phi_1$-strings for a larger $s$. 
On the other hand, for $s=2\eta_{1}^{-1}$, $R_4$ decreases compared to the case of $s=4\eta_{1}^{-1}$. 
By remembering that four $\phi_2$-strings
within a distance of $\order{\eta_1^{-1}}$ around a $\phi_1$-string should combine into
a compensated string, the decrease of $R_4$ 
for $s=2\eta_1^{-1}$ indicates the underestimate of the $\phi_2$-strings 
in a compensated string due to too small choice of $s$.
In our analysis,
we adopt $s=4\eta_{1}^{-1}$
as an optimal choice for $R_4$.

\bibliography{papers}

\end{document}